\documentclass[longauth]{aa} 

\usepackage{graphicx}
\usepackage{txfonts}
\begin{document} 

   \title{The GRAVITY Young Stellar Object survey}
   \subtitle{I. Probing the disks of Herbig Ae/Be stars in terrestrial orbits 
   \thanks{GTO programs with run ID: 0103.C-0347; 0102.C-0408; 0101.C-0311; 0100.C-0278; 099.C-0667}}
      \author{The GRAVITY Collaboration: K. Perraut\inst{1}\and L. Labadie\inst{2}\and B. Lazareff\inst{1}\and L. Klarmann\inst{3}\and D. Segura-Cox\inst{4}\and M. Benisty\inst{1,5}\and J. Bouvier\inst{1}\and W. Brandner\inst{3}\and A. Caratti o Garatti\inst{3,6}\and P. Caselli\inst{4}\and C. Dougados\inst{1}\and P. Garcia\inst{7,8,9}\and R. Garcia-Lopez\inst{3,6}\and S. Kendrew\inst{10,3}\and M. Koutoulaki\inst{3,6}\and P. Kervella\inst{11}\and C.-C. Lin\inst{3,12}\and J. Pineda\inst{4}\and J. Sanchez-Bermudez\inst{3,13}\and E. van Dishoeck\inst{4}\and R. Abuter\inst{14}\and A. Amorim\inst{7,15}\and J.-P. Berger\inst{1}\and H. Bonnet\inst{14}\and A. Buron\inst{4}\and F. Cantalloube\inst{3}\and Y. Cl\'enet \inst{11}\and V. Coud\'e du Foresto\inst{11}\and J. Dexter\inst{4}\and P.T. de Zeeuw\inst{4}\and G. Duvert\inst{1}\and A. Eckart\inst{2}\and F. Eisenhauer\inst{4}\and F. Eupen\inst{2}\and F. Gao\inst{4}\and E. Gendron\inst{11}\and R. Genzel\inst{4}\and S. Gillessen\inst{4}\and P. Gordo\inst{7,15}\and R. Grellmann\inst{2}\and X. Haubois\inst{9}\and F. Haussmann\inst{4}\and T. Henning\inst{3}\and S. Hippler\inst{3}\and M. Horrobin\inst{2}\and Z. Hubert\inst{1,11}\and L. Jocou\inst{1}\and S. Lacour\inst{11}\and J.-B. Le Bouquin\inst{1}\and P. L\'ena \inst{11}\and A. M\'erand\inst{14}\and T. Ott\inst{4}\and T. Paumard\inst{11}\and G. Perrin\inst{11}\and O. Pfuhl\inst{4}\and S. Rabien\inst{4}\and T. Ray\inst{6}\and C. Rau\inst{4}\and G. Rousset\inst{11}\and S. Scheithauer\inst{3}\and O. Straub\inst{4}\and C. Straubmeier\inst{2}\and E. Sturm\inst{4}\and F. Vincent\inst{11}\and I. Waisberg\inst{4}\and I. Wank\inst{2}\and F. Widmann\inst{4}\and E. Wieprecht\inst{4}\and M. Wiest\inst{2}\and E. Wiezorrek\inst{4}\and J. Woillez\inst{14}\and S. Yazici\inst{4,2}
          }

  \institute{
  Univ. Grenoble Alpes, CNRS, IPAG, 38000 Grenoble, France
\and
I. Physikalisches Institut, Universit\"at zu K\"oln, Z\"ulpicher Strasse 77, 50937, K\"oln, Germany
\and
Max Planck Institute for Astronomy, K\"onigstuhl 17,
69117 Heidelberg, Germany
\and
Max Planck Institute for Extraterrestrial Physics, Giessenbachstrasse, 85741 Garching bei M\"{u}nchen, Germany
\and
Unidad Mixta Internacional Franco-Chilena de Astronom\'ia (CNRS UMI 3386), Departamento de Astronom\'ia, Universidad de Chile, Camino El Observatorio 1515, Las Condes, Santiago, Chile     
\and
Dublin Institute for Advanced Studies, 31 Fitzwilliam Place, D02\,XF86 Dublin, Ireland
\and
CENTRA, Centro de Astrof\'{\i}sica e Gravita\c{c}\~{a}o, Instituto Superior T\'{e}cnico, Avenida Rovisco Pais 1, 1049 Lisboa, Portugal
\and
Universidade do Porto, Faculdade de Engenharia, Rua Dr. Roberto Frias, 4200-465 Porto, Portugal
\and 
European Southern Observatory, Casilla 19001, Santiago 19, Chile
\and
European Space Agency, Space Telescope Science Institute, 3700 San Martin Drive, Baltimore MD 21218, USA
\and
LESIA, Observatoire de Paris, PSL Research University, CNRS, Sorbonne Universit\'es, UPMC Univ. Paris 06, Univ. Paris Diderot, Sorbonne Paris Cit\'e, France
\and
Institute for Astronomy, University of Hawaii, 2680 Woodlawn Drive
96822 HI, USA
\and
Instituto de Astronom\'ia, Universidad Nacional Aut\'onoma de M\'exico, Apdo. Postal 70264, Ciudad de M\'exico 04510, Mexico
\and
European Southern Observatory, Karl-Schwarzschild-Str. 2, 85748
Garching, Germany
\and
Universidade de Lisboa - Faculdade de Ci\^encias, Campo Grande, 1749-016 Lisboa, Portugal\\
     Email: karine.perraut@univ-grenoble-alpes.fr   
}
   \date{Received ; accepted }

   \date{Received ; accepted }


\abstract
{The formation and the evolution of protoplanetary disks are important stages in the lifetime of stars. Terrestrial planets form or migrate within the innermost regions of these protoplanetary disks and so, the processes of disk evolution and planet formation are intrinsically linked. Studies of the dust distribution, composition, and evolution of these regions are  crucial to understanding planet formation. }
{We built a homogeneous observational dataset of Herbig Ae/Be disks with the aim of spatially resolving the sub au-scale region to gain a statistical understanding of their morphological and compositional properties, in addition to looking for correlations with stellar parameters, such as luminosity, mass, and age.}
{We observed 27 Herbig Ae/Be stars with the GRAVITY instrument installed at the combined focus of the Very Large Telescope Interferometer (VLTI) and operating in the near-infrared K-band, focused on the K-band thermal continuum, which corresponds to stellar flux reprocessed by the dust grains. Our sample covers a large range of effective temperatures, luminosities, masses, and ages for the intermediate-mass star population. The circumstellar disks in our sample also cover a range of various properties in terms of reprocessed flux, flared or flat morphology, and gaps. We developed semi-physical geometrical models to fit our interferometric data.}
{Our best-fit models correspond to smooth and wide rings that support previous findings in the H-band, implying that wedge-shaped rims at the dust sublimation edge are favored. The measured closure phases are generally non-null with a median value of $\sim$~10$^\circ$, indicating spatial asymmetries of the intensity distributions. Multi-size grain populations could explain the closure phase ranges below 20-25$^\circ$ but other scenarios should be invoked to explain the largest ones. Our measurements extend the Radius-Luminosity relation to $\sim$~10$^4$~L$_\odot$ luminosity values and confirm the significant spread around the mean relation observed by PIONIER in the H-band. Gapped sources exhibit a large N-to-K band size ratio and large values of this ratio are only observed for the members of our sample that would be older than 1\,Ma, less massive, and with lower luminosity. In the mass range of 2~$M_\odot$, we do observe a correlation in the increase of the relative age with the transition from group II to group I, and an increase of the N-to-K size ratio. However, the size of the current sample does not yet permit us to invoke a clear, universal evolution mechanism across the Herbig Ae/Be mass range. The measured locations of the K-band emission in our sample suggest that these disks might be structured by forming young planets, rather than by depletion due to EUV, FUV, and X-ray photo-evaporation.
}
 {}

    \authorrunning {K. Perraut et al.}
    \titlerunning{The GRAVITY YSO survey - Probing the disks of Herbig Ae/Be stars at terrestrial orbits}
   \maketitle
%

\begin{figure*}[t]
        \centering
        \includegraphics[width=15cm]{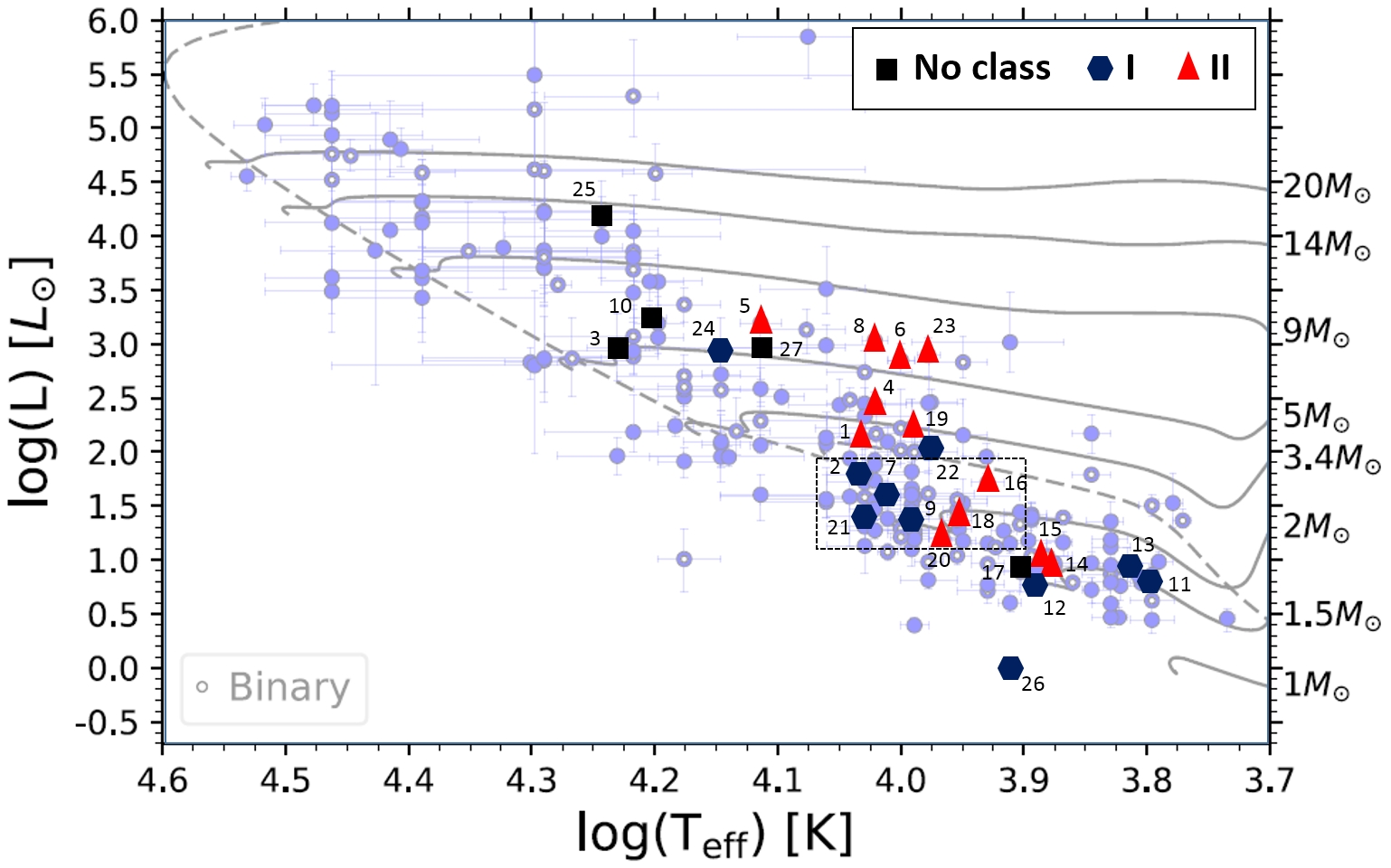}~ 
    \caption{GRAVITY (squares, diamonds, triangles) and Gaia (circles) observations \citep{Vioque2018A&A...620A.128V} of Herbig Ae/Be stars as put in the Hertzsprung-Russell diagram. The symbols code the Meeus groups \citep{Meeus2001}. The dashed line corresponds to the 2.5 Ma isochrone. The numbering refers to Table 1.} 
\label{fig:HR}
\end{figure*}

\section{Introduction}
An understanding how disks of gas and dust around young stars evolve and dissipate is essential to gaining a better understanding of planet formation theories. The morphology of protoplanetary disks provides information on their evolution and may relay evidence relevant to planet formation. In recent years, outer disk features have been probed in detail through scattered light imaging with instruments like SPHERE in the optical and near-infrared \citep{Beuzit2019arXiv190204080B} or through imaging at (sub-)millimetric wavelengths with ALMA \citep{ALMA2015ApJ...808L...1A}. ALMA has revealed structures in disks out to a few hundred astronomical units (au) \citep{Andrews2018Msngr.174...19A,Long_2018}, exhibiting multiple rings, gaps, and asymmetrical features, which possibly originate from dynamical interactions with giant planets \citep{vdM2016ApJ...832..178V,Zhang_2018,Lodato2019MNRAS.486..453L}. SPHERE images have also highlighted shadows, rings, spiral arms, warps, and gaps on the surface of the outer disks from 20 to 200~au \citep{Benisty2015A&A...578L...6B, deBoer2016A&A...595A.114D, Pohl2017A&A...605A..34P,Benisty2017A&A...597A..42B,Avenhaus2018ApJ...863...44A}. All of these observations clearly illustrate the complexity and  broad diversity of the morphology of the outer disks \citep{garufi}.

Currently, the level of detailed understanding with regard to the disk structure that is accessible with SPHERE or ALMA for the disks' outer region (i.e., $\sim$~20 to $\sim$~500~au) is lacking when we are looking at the innermost regions of a few au in size. However, the dynamical processes at work in these first central au might be just as complex and as diverse as what has been found in the outer regions. They most likely drive disk evolution in a key region where terrestrial planets form or migrate. The inner rim is the place where dust is processed thermally and whence it can be redistributed to the outer disk. There is evidence that the magneto-rotational instability activation in the innermost disk and its suppression in the dead zone are at the basis of the inside-out planet formation scenario \citep{Mohanty2018ApJ...861..144M}. In this context, dead zone and dust traps, resulting in the possible formation of local asymmetries, play a crucial role in the vortex generation and planetesimal formation \citep{Flock_2016,Flock_2017}. The interplay between these different aspects of the physics of disks is one of the important open question in the field, and a direct comparison between inner/outer disk phenomena is a possible way forward with regard to sharpening our view on young disk evolution.

The spectral energy distribution (SED) of the disk is too degenerate to allow the inner rim to be fully constrained, whereas zooming into the innermost regions of disks in the nearest star-forming regions (d$\sim$~120-130~pc) requires observations at milli-arcsecond (mas) resolution to reach the desired (sub-)au scales. As of a few recent decades, long-baseline infrared interferometry (LBI) has demonstrated such a unique ability, as, for example, with the IOTA interferometer \citep{MillanGabet} and the four-telescope H-band instrument PIONIER \citep{LeBouquin2011A&A...535A..67L} at the VLTI. Despite its poorer coverage of the spatial frequency plane in comparison to ALMA, LBI was able to bring new insights in the fundamental morphological properties of the inner disks of Young Stellar Objects (YSO; see \cite{Dullemond} and \cite{kraus2015} for a review). The size-luminosity relation adequately correlates the measured inner radii of disks with the luminosities of the central stars \citep{Monnier2005}, which is in strong agreement with the presence of a directly illuminated rim at the dust sublimation radius \citep{Natta2001}. To explain both photometric and interferometric measurements, and, in particular, the near-infrared (NIR) contribution of the rim to the Herbig NIR excess, several disk models have been proposed as vertically extended geometry \citep{Natta2001,Flock_2017}, disk wind components \citep{vinkovic,Bans}, or a tenuous dusty halo around the disk inner regions \citep{vinkovic2006,Maaskant2013A&A...555A..64M}. Regarding the inner rim structure, a wide range of shapes have been proposed: a puffed-up rim due to direct heating from the star, with a curved bright side of the rim linked to the gradient of evaporation temperature \citep{Natta2001,isella}; a wedge-shaped rim attributed to differences in cooling properties between grains of different sizes \citep{Tannirkulam};  a vertical structure of the rim when considering different grain
sizes and composition was self-consistently calculated by \cite{Kama2009}; see also Klarmann et al. subm.). The structure and the composition of these regions are still a matter of debate.

\begin{table*}[t]
\caption{Stellar properties from \cite{Vioque2018A&A...620A.128V}: distance $d$, effective temperature $T_{\rm eff}$, luminosity $L$, mass $M$, age, extinction $A_v$, and binarity flag $Bin$.
The last column gives the group of the Meeus classification \citep{Meeus2001,Juhasz2010} when available. Spectral type is from CDS-Simbad. Symbol $\star$ indicates the low-quality Herbig Ae/Be sample in \cite{Vioque2018A&A...620A.128V}.}
\centering
\vspace{0.1cm}
\begin{tabular}{c l c c c c c c c c c}
\hline
\hline
\# & Name & $d$ & $T_{\rm eff}$& $\log L$ & $M$ & Age & Spectral & A$_{\rm v}$ & $Bin$ & Meeus \\
 & & [pc] & [K] & [$L$ in L$_{\odot}$] & [M$_{\odot}$] & [Ma] & Type & [mag] & & classification \\
\hline
1 & HD 37806 & 428$^{+19}_{-16}$ & 10700$^{+1000}_{-700}$  &  2.17$^{+0.19}_{-0.14}$  & 3.11$^{+0.55}_{-0.33}$ & 1.56$^{+0.64}_{-0.60}$  &  B9 & 0.13$^{+0.19}_{-0.13}$ & Y & II \\ [1ex]
2 & HD 38120 & 405$^{+24}_{-20}$ & 10700$^{+800}_{-900}$ & 1.72$^{+0.31}_{-0.20}$ & 2.37$^{+0.43}_{-0.24}$ & 3$^{+14}_{-1}$ & B9 & 0.21$^{+0.50}_{-0.21}$ & & I\\[1ex]
3 & HD 45677   &     621$^{+41}_{-33}$  & 16500$^{+3000}_{-750}$        & 2.88$^{+0.32}_{-0.17}$   & 4.72$^{+1.19}_{-0.39}$ & 0.61$^{+3.77}_{-0.30}$ & B2& 0.57$^{+0.23}_{-0.15}$ & Y & \\[1ex]    
4 & HD 58647 &      319$^{+7.4}_{-6.8}$ & 10500$^{+200}_{-200}$ & 2.44$^{+0.11}_{-0.09}$        & 3.87$^{+0.33}_{-0.19}$  & 0.84$^{+0.12}_{-0.18}$ & B9& 0.37$^{+0.19}_{-0.12}$ & Y & II        \\[1ex]
5 & HD 85567 &       1023$^{+53}_{-45}$ & 13000$^{+500}_{-500}$ &        3.19$^{+0.10}_{-0.08}$  &  6.32$^{+0.53}_{-0.39}$       & 0.22$^{+0.05}_{-0.05}$ & B8 & 0.89$^{+0.03}_{-0.02}$ & Y & II \\[1ex]
6 & HD 95881  &      1168$^{+82}_{-66}$ & 10000$^{+250}_{-250}$ & 2.85$^{+0.10}_{-0.07}$ & 5.50$^{+0.50}_{-0.28}$ &      0.28$^{+0.05}_{-0.07}$ & B9.5 & 0.00$^{+0.05}_{-0.00}$ & & II\\  [1ex]  
7 & HD 97048   &     185$^{+2.2}_{2.1}$ & 10500$^{+500}_{-500}$ & 1.54$^{+0.07}_{-0.06}$ & 2.25$^{+0.11}_{-0.14}$        & 4.37$^{+1.11}_{-0.32}$ & A0 & 0.90$^{+0.05}_{-0.02}$ & & I\\ [1ex]  
8 & HD 98922   &     689$^{+28}_{-25}$  & 10500$^{+250}_{-250}$ & 3.03$^{+0.06}_{-0.05}$ &  6.17$^{+0.37}_{-0.31}$       & 0.20$^{+0.01}_{-0.04}$ & B9.5 & 0.09 $^{+0.01}_{-0.00}$ & Y & II\\ [1ex]
9 & HD 100546   &    110$^{+1}_{-1}$&    9750$^{+500}_{-500}$ &  1.37$^{+0.07}_{-0.05}$&      2.06$^{+0.10}_{-0.12}$&    5.5$^{+1.4}_{-0.77}$ & A0 & 0.00$^{+0.05}_{-0.00}$ & &I\\   [1ex]
10 & HD 114981 & 705$^{+57}_{-44}$ & 16000$^{+500}_{-500}$   & 3.24$^{+0.12}_{-0.09}$  & 6.09$^{+0.59}_{-0.34}$ & 0.28$^{+0.05}_{-0.07}$  & B5  & 0.00$^{+0.05}_{-0.00}$ & &  \\ [1ex]
11 & HD 135344B & 136$^{+2.4}_{-2.3}$ & 6375$^{+125}_{-125}$ & 0.79$^{+0.03}_{-0.04}$ & 1.43$^{+0.07}_{-0.07}$ & 8.93$^{+0.45}_{-0.91}$ & F8 & 0.23$^{+0.05}_{-0.06}$ & Y & I\\ [1ex]
12 & HD 139614 & 135$^{+1.6}_{-1.6}$ & 7750$^{+250}_{-250}$ & 0.77$^{+0.03}_{-0.01}$ & 1.48$^{+0.07}_{-0.07}$ & 14.5$^{+1.4}_{-3.6}$ & A9 & 0.00$^{+0.05}_{-0.00}$ & & I  \\ [1ex]
13 & HD 142527    &   157$^{+2.0}_{-1.9}$&       6500$^{+250}_{-250}$&   0.96$^{+0.03}_{-0.01}$ &     1.61$^{+0.12}_{-0.08}$ &   6.6$^{+0.3}_{-1.5}$ & F5 & 0.00$^{+0.05}_{-0.00}$ & Y & I\\[1ex]
14 & HD 142666    &   148$^{+2.0}_{-1.9}$&       7500$^{+250}_{-250}$ &  0.94$^{+0.04}_{-0.05}$  &   1.49$^{+0.08}_{-0.08}$ &    9.3$^{+0.8}_{-0.5}$ & A8 & 0.50$^{+0.08}_{-0.09}$ & & II\\[1ex]
15 & HD 144432     &  155$^{+2.4}_{-2.2}$&       7500$^{+250}_{-250}$ &  0.97$^{+0.04}_{-0.01}$ &     1.39$^{+0.07}_{-0.07}$ &   4.99$^{+0.25}_{-0.55}$ & A8  & 0.00$^{+0.06}_{-0.00}$ & Y & II\\[1ex]
16 & HD 144668      & 161$^{+3.1}_{-2.9}$ &      8500$^{+250}_{-250}$&   1.72$^{+0.05}_{-0.04}$ &    2.43$^{+0.12}_{-0.12}$&     2.73$^{+0.26}_{-0.35}$ & A3 & 0.33$^{+0.05}_{-0.04}$ & Y & II\\[1ex]
17 & HD 145718 & 153$^{+3.2}_{-3.0}$ & 8000$^{+250}_{-250}$ &        0.90$^{+0.05}_{-0.04}$ & 1.61$^{+0.08}_{-0.08}$ & 9.8$^{+2.8}_{-0.5}$ & A5 & 0.74$^{+0.06}_{-0.05}$ & Y & \\[1ex]
18 & HD 150193    &   151$^{+2.7}_{-2.5}$&       9000$^{+250}_{-250}$&   1.37$^{+0.04}_{-0.04}$  &   1.89$^{+0.09}_{-0.09}$ &    5.48$^{+0.44}_{-0.27}$ & A2 & 1.55$^{+0.02}_{-0.04}$ & Y & II\\[1ex]
19 & HD 158643     &  123$^{+8.2}_{-6.7}$&       9800$^{+900}_{-300}$&   2.22$^{+0.26}_{-0.07}$   &  3.35$^{+0.79}_{-0.22}$&     1.22$^{+0.29}_{-0.57}$ & A0 & 0.00$^{+0.34}_{-0.00}$ & & II \\[1ex]
20 & HD 163296      & 102$^{+2.0}_{-1.9}$&       9250$^{+250}_{-250}$&   1.2$^{+0.06}_{-0.03}$  &    1.83$^{+0.09}_{-0.09}$&     7.6$^{+1.1}_{-1.2}$ & A1 & 0.00$^{+0.05}_{-0.00}$ & & II\\[1ex]
21 & HD 169142       &114$^{+1.4}_{-1.3}$ &      10700$^{+800}_{-900}$  & 1.31$^{+0.12}_{-0.22}$  &    2.00$^{+0.13}_{-0.13}$     &   9$^{+11}_{-4}$ & F1 &  1.02$^{+0.12}_{-0.34}$ & & I\\[1ex]
22 & HD 179218  &     266$^{+5.6}_{-5.2}$ &      9500$^{+200}_{-200}$&   2.05$^{+0.09}_{-0.14}$   &   2.98$^{+0.18}_{-0.30}$&    1.66$^{+0.55}_{-0.26}$& A0 & 0.53$^{+0.12}_{-0.26}$ & Y & I\\[1ex]
23 & HD 190073 & 871$^{+96}_{-70}$&     9500$^{+200}_{-200}$& 2.90$^{+0.16}_{-0.20}$   & 5.89$^{+0.16}_{-0.20}$& 0.22$^{+0.11}_{-0.07}$ & A2 & 0.40$^{+0.12}_{-0.26}$ & &  II \\[1ex]
24 & HD 259431   &    721$^{+44}_{-37}$&         14000$^{+2125}_{-2900}$&        2.97$^{+0.27}_{-0.40}$   &    5.2$^{+1.8}_{-1.3}$ &     0.4$^{+0.5}_{-0.3}$ & B6 & 1.11$^{+0.21}_{-0.30}$ & Y & I\\[1ex]
25 & PDS 27 & 2550$^{+40}_{-310}$ & 17500$^{+3500}_{-3500}$ & 4.15$^{+0.37}_{-0.39}$ & 12.2$^{+5.5}_{-3.4}$ & 0.04$^{+0.07}_{-0.03}$ & B2 & 5.03$^{+0.13}_{-0.13}$ & & \\ [1ex]
26 & R CrA\,($\star$) & 95$^{+13}_{-9}$ & 8150$^{+180}_{-160}$ &  -0.06$^{+0.19}_{-0.19}$ & -- & -- & B5 & 2.13$^{+0.19}_{-0.16}$ & & I\\[1ex]
27 & V1818 Ori & 695$^{+82}_{-59}$ & 13000$^{+1000}_{-1500}$  & 2.96$^{+0.24}_{-0.29}$  & 5.3$^{+1.3}_{-1.1}$ & 0.37$^{+0.37}_{-0.19}$  & B7 & 3.78$^{+0.18}_{-0.24}$ & & \\ [1ex]
\hline
\end{tabular}
\label{tab:starproperties}
\end{table*}

With a particular focus on Herbig Ae/Be intermediate-mass objects, the H-band PIONIER Large Program (\cite{lazareff17}; L17) provides a detailed view on the inner rim region for a large number of stars ($\sim$~50) and it has shown that the inner rim, where dust is sublimated, appears smooth and wide, which is in agreement with pioneering CHARA observations \citep{Tannikurlam2008} and disk models with multi-grain populations. Because the disk/star contrast increases in the K-band, and the disk morphology in the continuum can be better compared to line emitting regions, we extend the H-band survey to the K-band to conduct a statistical study of the inner disk properties in this spectral range. This new survey is conducted in the context of the Guaranteed Time Observations (GTO) YSO program executed with the GRAVITY instrument \citep{GRAVITY} and which includes about a hundred young Herbig Ae/Be and T Tauri stars observed in the K-band continuum and across spectral lines such as the Hydrogen Br${\gamma}$ and the CO band heads.
Our targets span a wide range of masses (from 1 to 10 solar masses), ages (from 10$^4$ to 10$^7$~a), accretion rates (from 10$^{-6}$ to 10$^{-9}$~M$_\odot$/yr$^{-1}$), and disk morphologies (i.e., full and transitional disks), as we aim to explore the dependence of our results on the properties of the central star (spectral type, mass, luminosity, age) and those of the disk (reprocessed flux, presence of gaps, flared/flat morphology).

This paper reports on new K-band continuum observations with GRAVITY for 27 Herbig Ae/Be stars observed during the first two years of the GTO YSO program, with a focus on possible trends among different properties identified within the systems. The paper is organized as follows:  GRAVITY observations are described in Sect.~\ref{sect:obs}; we present our dataset in Sect.~\ref{sect:data}; our data analysis is detailed in Sect.~\ref{sect:drs}; the results are given in Sect.~\ref{sect:results} and discussed in Sect.~\ref{sect:disc}.

\begin{figure*}[t]
        \centering
        \includegraphics[width=17cm]{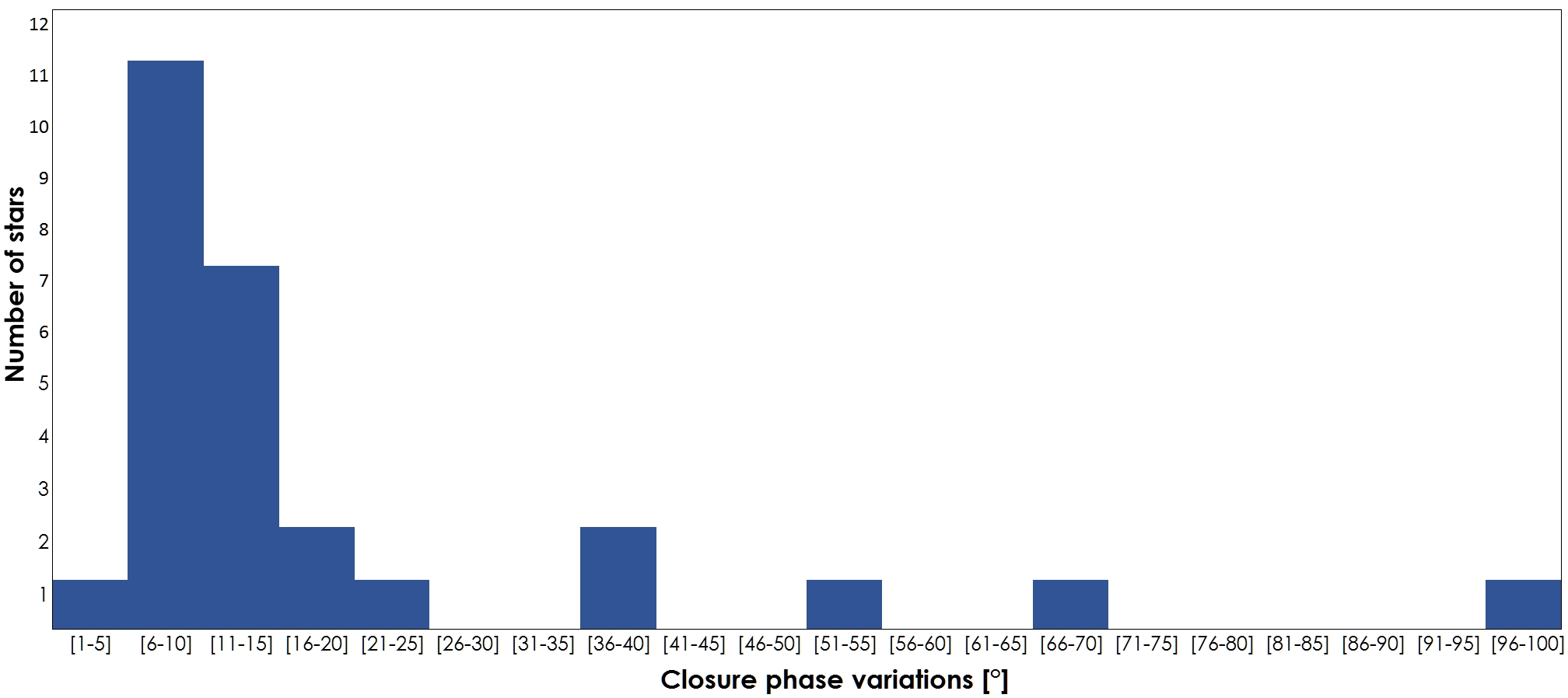}
    \caption{Histogram of  GRAVITY closure phase peak-to-peak variations over  spatial frequency range for our sample.}
    \label{fig:CP}
\end{figure*}

\section{Interferometric observations}
\label{sect:obs}

\subsection{Sample}

We observed a sample of 27 Herbig Ae/Be stars covering a broad range in luminosity (1-10$^4$~L$_\odot$), effective temperature (6375-17500~K), mass (1.4-12.2~M$_\odot$), and age (0.04-14.5~Ma), as detailed in Table~\ref{tab:starproperties} and Fig.~\ref{fig:HR}. Most of them are part of the PIONIER Large Program since they correspond to the brightest candidates, which will allow us to compare the disk morphology between the H- and K-bands. Since flat, flared, gapped, and ungapped disks are expected to have different signatures when observed at high angular resolution, we selected our targets to exhibit IR-excess and to include 10  group I objects, 12  group II objects, and 5 unclassified objects following the \citet{Meeus2001} classification (see Sect.~\ref{groups} for details).


\subsection{Observations}

All our targets were observed with the GRAVITY instrument, using the four 1.8-m Auxiliary Telescopes, with the exception of HD~100546, which was observed using the four 8-m Unit Telescopes. The interferometric signals were recorded on six baselines simultaneously on the fringe tracker (FT) and the science channel (SC) detectors. The SC records the interferometric observables at high spectral resolution ($\cal{R}$~$\sim$~4000) over the whole K-band with individual integration times of, typically, 10~s to 30~s. The FT records at low spectral resolution (five spectral channels over the K-band) at frame rates ranging from $\approx$300  to $\approx$~900~Hz \citep{Lacour2019A&A...624A..99L}. Working at such a high speed allows for the atmospheric effects to be frozen. Each observation file corresponds to five minutes on the object. To calibrate the atmospheric transfer function, we interleaved our target observations with observations of interferometric calibrators that are established as single stars, bright, small enough, and close to the target. A detailed log of the observations is presented in Table A.1. 

\subsection{Data reduction}

All the data were reduced using the GRAVITY data reduction pipeline \citep{DRS}. In this paper, we focus on the FT observations for probing the inner dust rim in the K-band continuum as the turbulence effects are much lower at the speed of the FT. The SC observations at a spectral resolution of about 4000 will be used to constrain the Br$\gamma$ line emitting regions (Garcia-Lopez et al., in prep). For each file, we obtained six squared visibilities and four closure phases for six spectral channels. In the following, we discard the first spectral channel that might be affected by the metrology laser working at 1.908~$\mu$m and by the strong absorption lines of the atmospheric transmission.

\section{Dataset}
\label{sect:data}

 All the calibrated data are displayed in Appendix B. Our GRAVITY observations span a spatial frequency range between 5~M$\lambda$ and 60~M$\lambda$. Our maximum angular resolution of $\lambda$/2B is of about 1.7 millisecond of arc (mas) for a longest baseline B of 130~m, which corresponds to about 0.24~au at a distance of 140~pc. Except for 5 objects that are partially resolved (HD~85567, HD~114981, HD~158643, HD~259431, and PDS~27) with minimum squared visibilities larger than 0.4, our squared visibility measurements go below 0.4. A few objects are fully resolved and their visibilities are close to 0 at the longest baselines (HD~37806, HD~45677, HD~98922, HD~100546, HD~190073, and R~CrA). For each object, the visibility variation with the spatial frequency allows the extent of the environment to be determined through model fitting (Sect.~\ref{sect:drs}).

The closure phases are related to the asymmetry of the environment \citep{Haniff2007}: the closure phase is null for a centro-symmetric target; in the case of asymmetrical objects, the closure phase signal increases with spatial resolution up to the point where the spatial scale of the asymmetry is resolved; beyond this point, the closure phase signal decreases. In looking at our sample, the histogram of closure phase variations over the spatial frequency range, i.e. the peak-to-peak difference between the minimum and the maximum closure phases, shows that only one star among 27 has a variation smaller than 5$^\circ$. About half of our targets (12 among 27) have closure phase variations smaller than 10$^\circ$; about a third (10 among 27) has a closure phase variation ranging between 11$^\circ$ and 25$^\circ$; and the five other targets have strong closure variations that exceed 30$^\circ$ (Fig.~\ref{fig:CP}). As expected, the strong closure phase signals are observed for well-resolved targets (HD~45677, HD~98922, HD~144432, HD~144668, HD~179218, and R~CrA). Conversely, several targets are well-resolved with visibilities squared below 0.2 and do not exhibit closure phase variations higher than 15$^\circ$ (HD~37806, HD~58647, HD~97048, HD~135344B, HD~142527, HD~150193, HD~163296, SAO~206462).

\section{Geometric modelling}
\label{sect:drs}

We use the same analysis tools as in the PIONIER survey and developed similar geometric models as described in L17. In the following section, we only recall the main steps involved in the fitting method.

\subsection{Model visibility}

At near-infrared wavelengths, the disk emission essentially comes from a compact region near the sublimation radius, which justifies the adoption of simple geometric models capable of capturing the general morphological properties of the environment. This would be less true at mid-infrared wavelengths, where the intensity distribution is spatially more extended. Accordingly, our model includes a point-like central star and a circumstellar environment. The complex visibility at the spatial frequencies ($u$, $v$) and at the wavelength $\lambda$ is, therefore, described by a linear combination of the two components as follows:
\begin{equation}
    V (u,v,\lambda) = f_s (\lambda) + f_c (\lambda) V_c (u,v),
\end{equation}
where the star visibility equals 1 since it is assumed to be unresolved (even for the closest star forming regions, the interferometric angular resolution is not high enough to resolve the stellar photosphere). $V_c$ stands for the visibility of the circumstellar environment, $f_s$ and $f_c$ for the fractional flux contributions of the star and the circumstellar environment, respectively ($f_s$ + $f_c$~=~1). With regard to targets for which the scattered light amount is significant (mainly for transitional disks; see L17), an additional flux contribution (called {\it halo} and characterized by a flux ratio $f_h$) can be considered. The halo is assumed to be much more extended than the angular resolution of the interferometer. Its visibility is, thus, null. Because the halo contributes to the fractional flux (i.e., $f_s$ + $f_c$ + $f_h$~=~1), its net effect is that for very short baselines, the visibility is smaller than 1. At very long baselines, the environment might be completely resolved and its visibility $V_c$ reaches zero. In that case, the visibility curves display a plateau, which corresponds to the stellar flux contribution $f_s$.

\subsection{Flux ratio determination}
\label{sect:sed}

To determine the flux ratio between the star and the environment for the Herbig Ae/Be stars, we use the same photometry datasets as the PIONIER LP. For each target, we fit the SED under the same assumptions as L17: we perform a least-square fit of the observed fluxes in the U, B, V, R, I, J, H, K Johnson-Cousins bands; the stellar photosphere is represented by the relative fluxes in each Johnson-Cousins band for the relevant spectral type, while the dust emission is modeled as a single temperature blackbody. The fitted parameters are the absorption-free stellar flux in V band, the absorption-free dust flux in the K-band, the extinction Av, and the blackbody temperature of the dust component. The K-band environment contribution is expected to be higher than in the H-band one. Being able to accurately determine the stellar flux with only our GRAVITY observations would require to reach the plateau when the circumstellar environment is fully resolved, which is only the case for a few stars. The SED fit provides us with the fractional flux $f_c$ of the circumstellar environment at the central wavelength of the K-band; that value, together with its error bar, acts as a constraint in the interferometric fit, but it is kept as a free parameter during the fitting process for visibility (see Table 2).

\subsection{Spectral dependence}

As described in Sect.~\ref{sect:data}, our GRAVITY FT data contain six visibilities and four closure phases for each of the five spectral channels. We use this spectral information to derive the spectral index of the circumstellar environment and model our visibility with the following formula

\begin{equation}
    V (u,v,\lambda) = \frac{f_s (\lambda_0/\lambda)^{k_s} + f_c (\lambda_0/\lambda)^{k_c} V_c (u,v)}{(f_s + f_h) (\lambda_0/\lambda)^{k_s} + f_c (\lambda_0/\lambda)^{k_c}},
\end{equation}
with $\lambda_0$~=~2.15~$\mu$m the wavelength of the central spectral channel of the FT, and $k$ the spectral index defined by
\begin{equation}
    k = \frac{d \log F_\nu}{d \log \nu},
\end{equation}
where $k_s$ and $k_c$ are the spectral indices of the star and the circumstellar environment, respectively. 

For this interferometric data fit, the star photosphere is approximated as a blackbody at the stellar effective temperature over the K-band. While differences between a blackbody and a Kurucz model are large in the ultraviolet domain, in the near-infrared range the differences are small for the effective temperature range of the star between 10000~K to 20000~K: the photospheric absorption lines across the K-band are scarce and shallow, the spectral slope of the continuum is almost identical to the blackbody, the absolute flux levels can differ by 15-30\% but only the slope is relevant when fitting our visibility data. Regarding the spectral indices, $k_s$ is derived from the effective temperature of the star (see Sect.~\ref{sect:sed}) using Table 5 of \citet{pecaut13}. $k_c$ is an additional free parameter for our geometric models and it can be converted into an equivalent temperature of the dust $T_d$, assuming a blackbody emission.

\subsection{Fitting processes}

The interval between different observing epochs is shorter than 1 month for most of our targets, and about two months for three targets (HD~144668, HD~150198, and HD~158643). For HD~259431, the two observing sequences are spaced out by close to a year but our fit is entirely dominated by the 12 observations obtained in 2018. For HD~144668, for which we have seven and nine observing files, respectively, we have checked that the fits lead to consistent sizes and orientation when considering each dataset independently. Since there is, so far, no observational evidence of temporal variability at the spatial resolution of our dataset, we have combined the different epochs -- when available -- for each object in our survey prior to the fitting process. This provides us with a statistically robust dataset against the number of free parameters, assuming that the near-infrared emission and the disk structures do not vary. 

For the purposes of consistency, the observational data are fitted independently following two different numerical approaches: one based on Markov chain Monte Carlo (MCMC) and one based on a Levenberg-Marquardt algorithm. Since both methods provide results consistent within 3-$\sigma$, in this paper, we only present those of the MCMC approach.

We use the same tools as those developed for the PIONIER Large Program (see L17 for details) based on two steps of the Shuffled Complex Evolution (SCE) algorithm \citep{Duan}, and then one final MCMC \citep{Haario}. The circumstellar environment can be modeled either as an ellipsoid or as a ring. The radial brightness distribution of both models varies from a Gaussian profile to a Lorentzian one (through the $\rm Lor$ parameter; see Appendix D). With such models, the brightness distribution is centro-symmetric and the resulting closure phase is null. The ring model (with a half-flux radius of $a_r$) is based on a wireframe image convolved by an ellipsoid kernel of semi-major axis $a_k$ and the same axial ratio $\cos i$ as the wireframe. This permits the description of several kinds of rings, from a thin ring (for very small values of $a_k$ with respect to $a_r$) to a very wide ring tending to ellipsoids (for large values of $a_k$ with respect to $a_r$). In addition, the ring model can include an angular first-order modulation involving sine terms to mimic an inner rim of a disk that is viewed as inclined and that appears as an ellipse with a brighter far side. Such skewed ring models were first introduced to attempt to account for non-zero closure phases in the first IOTA observations \citep{Monnier2006}. Indeed, for such a model, the brightness distribution is not centro-symmetric and non-zero closure phases can be computed.

Since the error bars on our observables can be underestimated and/or correlated, we impose floor values to the error estimates (from the reduction pipeline) of respectively 2\% for the squared visibilities and 1$^\circ$ for the closure phases.
Even so, the reduced{\em } $\chi^2_r$ is generally larger than unity; assuming this results from under-estimated error estimates for the observed quantities, we rescale the $\chi^2$ supplied to the SCE and MCMC algorithms such that $\chi^2_r\approx 1$, correspondingly increasing the error estimates for the derived model parameters. The reader can refer to Sect. 3.3 of L17 for details. While this procedure addresses the issue of optimistic error bars on the observables, there are cases where the issue is clearly not the value of the model parameters, but the model itself. \\

The free parameters of our models are gathered in Table~2. The ellipsoid model has 7 free parameters; the ring model without azimuthal modulation has 8 and the ring model with a first mode azimuthal modulation has 10.

\begin{figure*}[t]
        \centering
        \includegraphics[width=9.3cm]{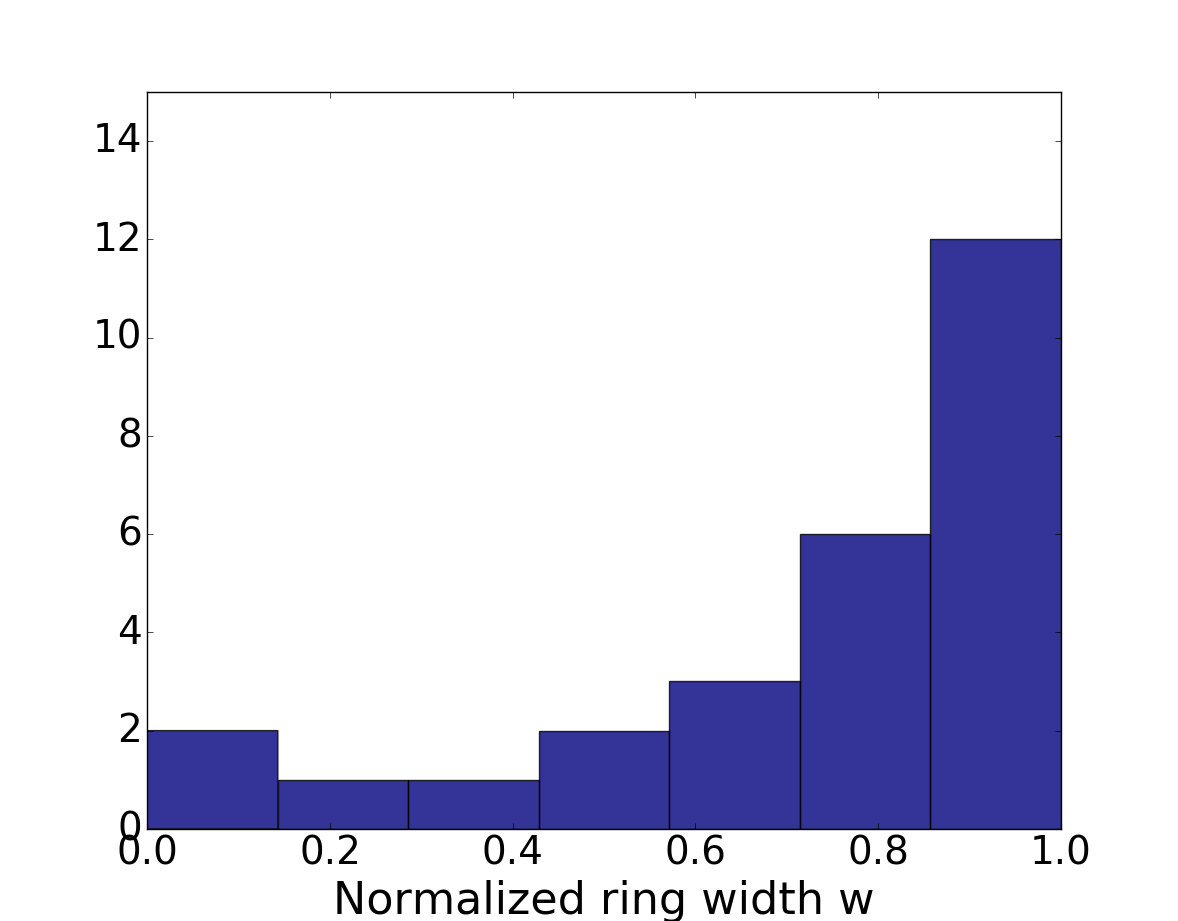}
    \includegraphics[width=9cm]{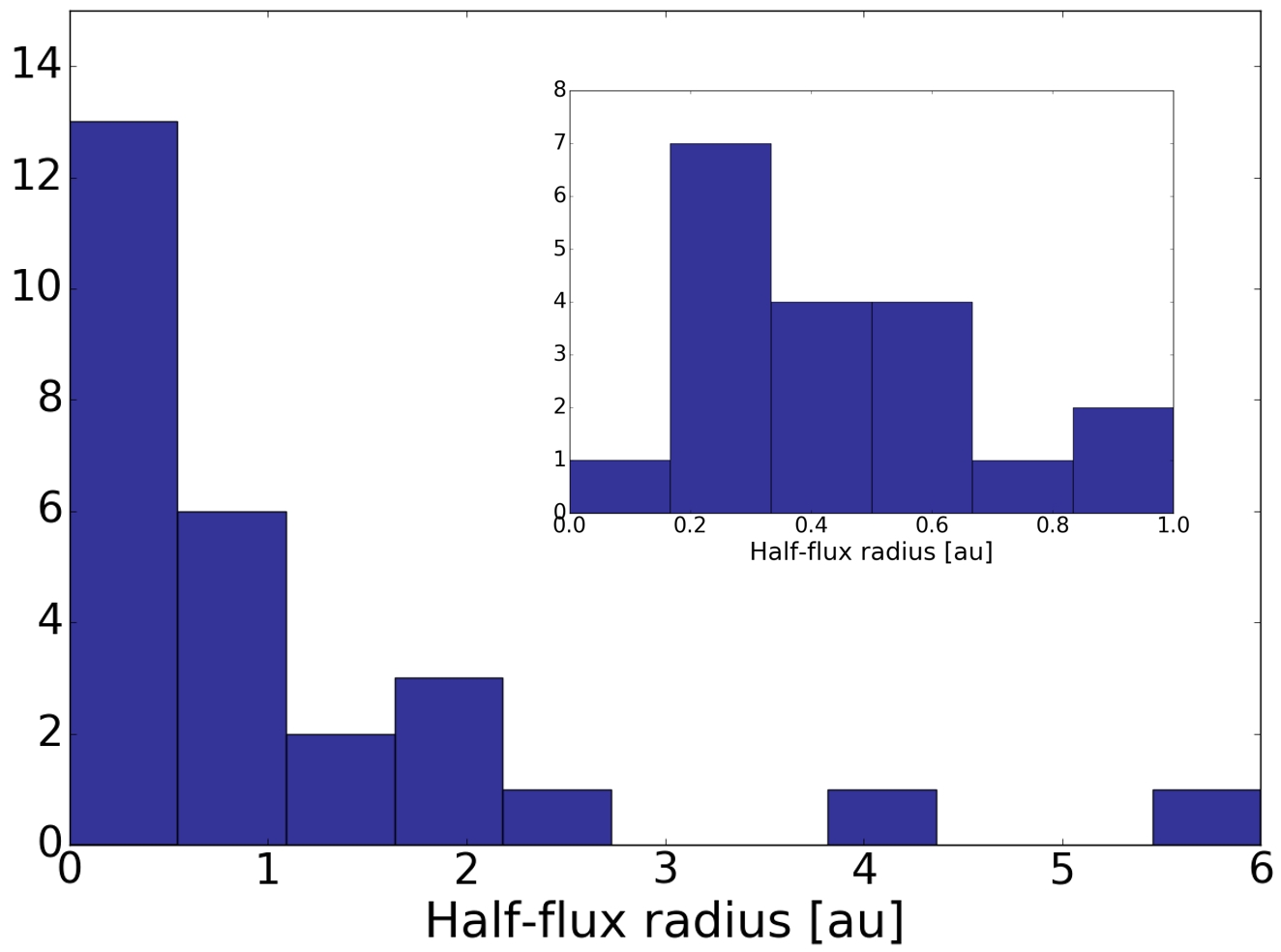}
    \caption{{\bf Left.} Histogram of  normalized width $w$ for ring model of our GRAVITY observations. {\bf Right} Half-flux radius {\bf $a$} derived from our GRAVITY measurements and  Gaia DR2 distances. Insert is a zoom on the shortest radii.}
    \label{fig:W}
\end{figure*}

\begin{table}[h]
\label{tab:param}
\caption{Free parameters of our geometric models.}
\vspace{0.1cm}
\begin{tabular}{c p{6.5cm}}
\hline \hline
$f_c$ & Flux contribution of circumstellar environment \\
$f_h$ & Flux contribution of  halo (if any) \\
$\cos i$ & Cosine of inclination\\
$PA$ & Position Angle of major axis from North to East \\
$k_c$ & Spectral index of  circumstellar environment \\
\hline
{\bf Ellipsoid} & \\
$\rm Lor$ & Weighting for radial brightness distribution$^{(a)}$\\
$\log a$ & Log of half-flux semi-major axis (with $a$ in mas)\\
\hline
{\bf Ring} & \\
$\rm Lor$ & Weighting for radial brightness distribution of  convolution kernel$^{(a)}$\\
$\log a$ & Log of half-flux semi-major axis (with $a$ in mas)\\
$\log (a_k/a_r)$ & Log of ratio between angular radius of kernel and of ring\\
$c_1$, $s_1$ & Cosine and sine angular modulation of 1st order\\
\hline
\end{tabular}
{\footnotesize (a) $\rm Lor$ ranges between 0 and 1; $\rm Lor$~=~0 for a  Gaussian radial distribution and $\rm Lor$~=~1 for a Lorentzian one. See Appendix D.}
\end{table}

\begin{table*}[t]
\centering
\label{tab:ring}
\caption{Best-fit parameters for  ring models with 1-$\sigma$ error bars as defined in Table 2. Closure phases are not included in the $\chi^2$ value. Column 1 corresponds the numbering of Table 1. Column 8 provides the half-flux diameter ($FWHM~=~2~a$) and column 10 provides the dust temperature at 2.2~$\mu$m derived from the spectral index $k_c$ (see Sect.~5.4).}
{\small
\begin{tabular}{c c c c c c c c c c c}
\hline \hline
 &Object & $f_c$ & $f_h$ & $\cos i$ & $PA$ & $\rm Lor$ & $FWHM$ & $w~=~a_k/a$ & $T_{\rm dust}$ & $\chi_r^2$\\
& -- & -- & -- & -- & [$^\circ$] & -- & [mas]  & -- & [K] & -- \\
\hline
1 & HD~37806 & 0.85~$\pm$~0.01 & 0.00$\pm$~0.00 & 0.50$\pm$~0.04 & 53~$\pm$~1 & 0.96~$\pm$~0.09 & 4.08~$\pm$~0.10 & 0.69~$\pm$~0.06 & 1440~$\pm$~90 & 0.69\\
2 & HD~38120 & 0.64~$\pm$~0.01 & 0.00$\pm$~0.01 & 0.66$\pm$~0.02 & 165~$\pm$~2 & 0.48~$\pm$~0.13 & 6.47~$\pm$~0.15 & 0.74~$\pm$~0.23 & 1480~$\pm$~90 & 5.37\\
3 & HD~45677 & 0.89~$\pm$~0.01 & 0.0~$\pm$~0.0 & 0.51~$\pm$~0.01 & 66~$\pm$~1 & 0.98~$\pm$~0.04 & 19.1~$\pm$~0.40 &  0.74~$\pm$~0.02 & 1120~$\pm$~70 & 0.23\\
4 & HD~58647 & 0.66~$\pm$~0.01 & 0.00~$\pm$~0.00 & 0.45~$\pm$~0.01 & 16~$\pm$~1 & 0.98~$\pm$~0.04 & 3.99~$\pm$~0.09 & 0.45~$\pm$~0.03 & 1140~$\pm$~50 & 0.76\\
5 & HD~85567 & 0.89~$\pm$~0.02 & 0.06~$\pm$~0.01 & 0.93~$\pm$~0.01 & 92~$\pm$~7 & 0.15~$\pm$~0.06 & 1.17~$\pm$~0.02 & 0.28~$\pm$~0.24 & 1680~$\pm$~180 & 3.02\\
6 & HD~95881 & 0.88~$\pm$~0.01 & 0.00~$\pm$~0.03 & 0.60~$\pm$~0.03 & 161~$\pm$~5 & 0.68~$\pm$~0.16 & 2.89~$\pm$~0.20 &  0.98~$\pm$~0.02 & 1590~$\pm$~60 & 15.60\\
7 & HD~97048 & 0.73~$\pm$~0.02 & 0.01~$\pm$~0.01 & 0.66~$\pm$~0.01 & 179~$\pm$~1 & 1.00~$\pm$~0.01 & 4.38~$\pm$~0.10 & 0.83~$\pm$~0.02 & 1295~$\pm$~60 & 2.60\\
8 & HD~98922 & 0.83~$\pm$~0.01 & 0.0~$\pm$~0.0 & 0.77~$\pm$~0.01 & 119~$\pm$~2 & 1.00~$\pm$~0.01 & 4.69~$\pm$~0.11 &  0.83~$\pm$~0.01 & 1410~$\pm$~60 & 0.72\\
9 & HD~100546 & 0.63~$\pm$~0.01 & 0.00~$\pm$~0.01 & 0.64~$\pm$~0.01 & 146~$\pm$~1 & 0.99~$\pm$~0.02 & 5.02~$\pm$~0.12 &  0.46~$\pm$~0.02 & 1360~$\pm$~60 & 0.09\\
10 & HD~114981 & 0.14~$\pm$~0.01 & 0.00$\pm$~0.01 & 0.29$\pm$~0.07 & 167~$\pm$~4 & 0.04~$\pm$~0.33 & 12.1~$\pm$~0.5 & 0.10~$\pm$~0.03 & 1380~$\pm$~110 & 29.5\\
11 & HD 135344B & 0.56~$\pm$~0.01 & 0.00$\pm$~0.01 & 0.76$\pm$~0.01 & 30~$\pm$~2 & 0.98~$\pm$~0.08 & 2.96~$\pm$~0.07 & 0.38~$\pm$~0.04 & 1620~$\pm$~60 & 2.60\\
12 & HD~139614 & 0.58~$\pm$~0.01 & 0.00$\pm$~0.00 & 0.75$\pm$~0.03 & 155~$\pm$~5 & 0.97~$\pm$~0.10 & 9.57~$\pm$~0.43 & 0.87~$\pm$~0.30 & 1140~$\pm$~40 & 0.53\\
13 & HD~142527 & 0.66~$\pm$~0.02 & 0.0~$\pm$~0.0 & 0.91~$\pm$~0.02 & 14~$\pm$~4 & 1.00~$\pm$~0.04 & 2.52~$\pm$~0.06 & 0.70~$\pm$~0.03 & 1480~$\pm$~60 & 0.94\\
14 & HD~142666 & 0.74~$\pm$~0.03 & 0.21~$\pm$~0.03 & 0.56~$\pm$~0.04 & 162~$\pm$~3 & 0.99~$\pm$~0.04 & 1.45~$\pm$~0.26 & 0.99~$\pm$~0.04 & 1600~$\pm$~110 & 18.2\\
15 & HD~144432 & 0.67~$\pm$~0.02 & 0.00~$\pm$~0.00 & 0.87~$\pm$~0.01 & 76~$\pm$~3 & 0.96~$\pm$~0.14 & 2.76~$\pm$~0.07 & 0.58~$\pm$~0.06 & 1570~$\pm$~80 & 2.07\\
16 & HD~144668 & 0.78~$\pm$~0.01 & 0.0~$\pm$~0.0 & 0.56~$\pm$~0.01 & 123~$\pm$~1 & 0.74~$\pm$~0.07 & 4.08~$\pm$~0.10 &  0.72~$\pm$~0.05 & 1730~$\pm$~60 & 1.09\\
17 & HD~145718 & 0.48~$\pm$~0.01 & 0.00$\pm$~0.00 & 0.31$\pm$~0.02 & 2~$\pm$~2 & 0.92~$\pm$~0.07 & 9.14~$\pm$~0.66 & 1.00~$\pm$~0.01 & 1470~$\pm$~60 & 5.07\\
18 & HD~150193 & 0.87~$\pm$~0.01 & 0.00~$\pm$~0.01 & 0.68~$\pm$~0.03 & 176~$\pm$~3 & 0.90~$\pm$~0.05 & 5.26~$\pm$~0.12 & 1.00~$\pm$~0.01 & 1965~$\pm$~80 & 4.27\\
19 & HD~158643 & 0.25~$\pm$~0.01 & 0.0~$\pm$~0.0 & 0.46~$\pm$~0.01 & 116~$\pm$~3 & 0.64~$\pm$~0.18 & 6.18~$\pm$~0.14 &  0.97~$\pm$~0.22 & 1150~$\pm$~30 & 1.55\\
20 & HD~163296 & 0.81~$\pm$~0.01 & 0.0~$\pm$~0.0 & 0.64~$\pm$~0.01 & 133~$\pm$~1 & 0.73~$\pm$~0.08 & 5.90~$\pm$~0.14 &  1.00~$\pm$~0.02 & 1560~$\pm$~70 & 2.17\\
21 & HD~169142 & 0.40~$\pm$~0.05 & 0.09~$\pm$~0.05 & 0.8~$\pm$~0.2 & 40~$\pm$~30 & 1.00~$\pm$~0.04 & 5.8~$\pm$~1.3 & 0.93$\pm$~0.13 & 1300~$\pm$~80 & 2.6\\
22 & HD~179218 & 0.64~$\pm$~0.01 & 0.0~$\pm$~0.0 & 0.59~$\pm$~0.09 & 68~$\pm$~10 & 0.64~$\pm$~0.18 & 17.0~$\pm$~2.1 & 1.00~$\pm$~0.13 & 1170~$\pm$~50 & 3.74\\
23 & HD~190073 & 0.80~$\pm$~0.01 & 0.00~$\pm$~0.01 & 0.93~$\pm$~0.01 & 29~$\pm$~6 & 0.95~$\pm$~0.16 & 4.08~$\pm$~0.10 & 0.14~$\pm$~0.03 & 1520~$\pm$~80 & 1.22\\
24 & HD~259431 & 0.84~$\pm$~0.02 & 0.00~$\pm$~0.01 & 0.83~$\pm$~0.01 & 49~$\pm$~3 & 1.00~$\pm$~0.01 & 1.00~$\pm$~0.03  & 1.00~$\pm$~0.01 & 1710~$\pm$~120 & 1.75\\
25 & PDS 27 & 0.80~$\pm$~0.10 & 0.10$\pm$~0.02 & 0.95$\pm$~0.04 & 158~$\pm$~28 & 0.18~$\pm$~0.10 & 1.66~$\pm$~0.07 & 0.85~$\pm$~0.13 & 1690~$\pm$~90 & 3.44\\
26 & R CrA & 0.98~$\pm$~0.01 & 0.00~$\pm$~0.01 & 0.59~$\pm$~0.05 & 4~$\pm$~3 & 1.00~$\pm$~0.02 & 11.2~$\pm$~0.6  & 1.00~$\pm$~0.01 & 1340~$\pm$~80 & 0.40\\
27 & V1818 Ori & 0.63~$\pm$~0.08 & 0.10$\pm$~0.02 & 0.89$\pm$~0.01 & 131~$\pm$~3 & 0.10~$\pm$~0.14 & 2.70~$\pm$~0.70 & 0.97~$\pm$~0.43 & 1440~$\pm$~190 & 0.73\\
\hline
\end{tabular}
}
\end{table*}
\section{Results}
\label{sect:results}

As the closure phase signals are generally small, we first focused on centro-symmetric models to derive characteristic sizes for the environment. For all targets, we tested three different models: (a) an ellipsoid leaving $\rm Lor$ free, (b) an ellipsoid with $\rm Lor$ fixed to zero for a Gaussian distribution, (c) a ring leaving $\rm Lor$ free. Finally we tested models without halo by fixing $f_h$ to zero. We compared the models by comparing the $\chi^2_r$ values.
 
\subsection{General findings}
\label{sect:general}

The best agreements are obtained for models with free $\rm Lor$. Gaussian models with halo generally lead to slightly worse $\chi^2_r$ while Gaussian models without halo exhibit much worse  $\chi^2_r$ than the other models and, thus, they are not considered further. For comparison, Table~3 summarizes the parameters for the best ring models with free $\rm Lor$ and Table~C.1 summarizes the parameters for the best Gaussian ellipsoid models. For 21 targets (78\%), the best-fit angular sizes are similar among all models. For the other 6 targets (HD~37806, HD~97048, HD~98922, HD~114981, HD~139614, R~CrA), since the models with the free intensity distributions are better than the Gaussian ones, we discuss in the following the best-fit parameters of Table~3. For all models, the inclinations $i$ and the position angles $PA$ are well-constrained and do not depend on the models within an accuracy of $\pm$~5$^\circ$.

Our models do not fit the data properly for three targets (HD~95881, HD~114981, and HD~142666) since the best models have $\chi^2_r$ larger than 15. We retain them in our sample despite the poor fits.

\subsection{Halo contribution}

When fitting the environment with a Gaussian intensity distribution, 70\% of the models require a halo contribution (Table~C.1.): the halo contribution reaches 10\% or more for 8 objects (30\%), varies between 5\% and 9\% for 5 objects (19\%), and is below 5\% for 6 objects (22\%). In contrast, when fitting with a free $\rm Lor$, the best fits generally converge towards models with null halo contribution, except for HD~85567, HD~97048, HD~142666, HD~169142, PDS~27, and V1818~Ori (Table~3).

The halo contribution determination requires very short baselines, as reported in \citet{Setterholm2018ApJ...869..164S}. These are generally not covered in our (u, v) planes but the halo contribution in the H-band as constrained by the PIONIER observations is null or below 6\% for all of our targets except HD~100546 (11\%) and HD~169142 (8\%) (see L17). This halo contribution is expected to be no larger in the K-band than in the H-band when interpreted as scattered light \citep{Pinte2008}.

\subsection{Inner disk morphology}

Using our $\rm FWHM$ values and  Gaia distances (Table~\ref{tab:starproperties}), we derived the half-flux radii (Fig.~\ref{fig:W}-right) of the environments; these range from 0.1 to about 6 astronomical units (au) with a median of 0.6 au.

\begin{figure*}[t]
        \centering
        \vspace{0.5cm}
        \includegraphics[width=16cm]{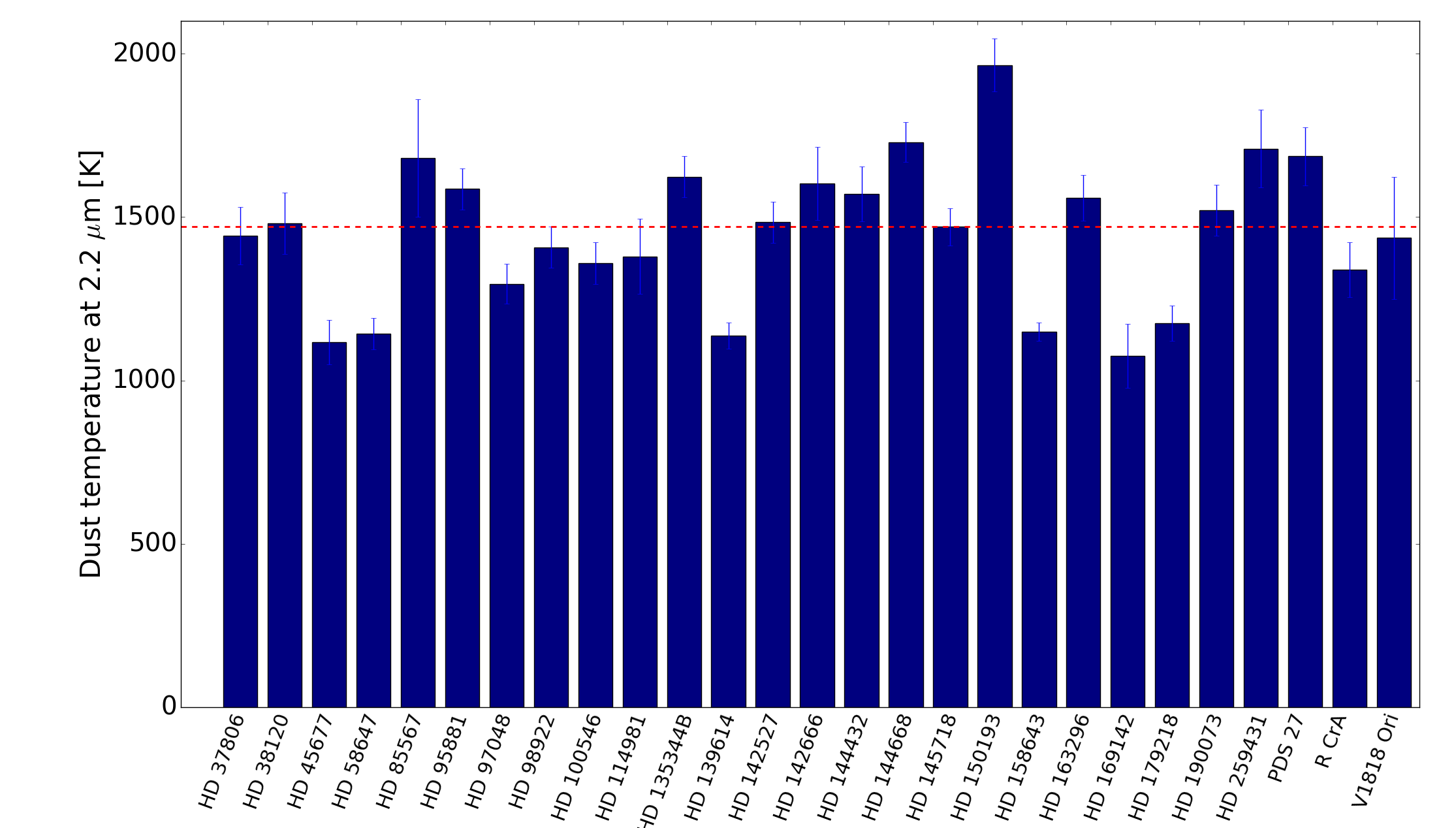}
    \caption{Bar plot of temperatures of environment dust at 2.2~$\mu$m, assuming a blackbody emission. Dashed line shows the median dust temperature. 
    }
    \label{fig:TdustK}
\end{figure*}

Comparing the results obtained for the ellipsoid and the ring models shows that rings generally have a lower $\chi^2_r$. The best fits for the ring are, most often, wide rings, that is, with a ratio of width to half-light radius $w = a_k/a$ of the order of unity. Considering all the targets, the median of the relative width $w$ equals 0.83 (Fig.~\ref{fig:W}-left) and the error bars on ring width are generally large. Thus, we cannot derive an accurate width of the ring and observe an inner cavity, meaning one without any dust inside a sharp inner rim. This result is in agreement with previous PIONIER findings, where only a few objects show inner cavities, even when higher spatial frequencies are probed.
  
Regarding the disk inclination, the fitted values of $\cos i$ range from about 0.3 and 0.95 (i.e., $i$ between $\sim$~20$^\circ$ and $\sim$~70$^\circ$), confirming that pole-on assumption ($i$~=~0$^\circ$) for characteristic size determination is not realistic for Herbig Ae/Be stars, despite the low optical extinction usually invoked.

\subsection{Dust temperature at 2.2~$\mu$m}
\label{spectral}

Using the spectral dependence of our interferometric data and assuming the spectral index of the stellar photosphere from the star spectral type, we fitted the spectral index of the environment and converted it into dust temperature at 2.2 $\mu$m under the following assumptions: (a) the circumstellar contribution is dominated by the dust radiation while the circumstellar environment produces an emission like a black (gray) body, and (b) there is no non-photospheric unresolved emission. A median dust temperature of $\sim$1470\,K with a standard deviation of $\sim$200\,K is measured (Fig.~\ref{fig:TdustK}). This is comparable with the sublimation temperature of silicates (1500~K) and of carbon (1800~K) compositions. It could be compared with the PIONIER findings that favor sublimation temperatures which are definitely larger than 1500~K for the dust responsible for the H-band emission (see Sect.~6.2).


\subsection{Which asymmetry do our non-null closure phase signals probe?}

Most objects of our sample exhibit non-zero closure phase differences, indicative of an asymmetry of the environment.
As seen in Fig.~\ref{fig:CP}, while about 70\% of our targets have closure phase variations over the spatial frequency range smaller than 15$^\circ$, a few stars exhibit strong closure phase variations (i.e., larger than 25$^\circ$). This is the case for HD~45677, HD~98922, HD~144432, HD~144668, HD~179218, and R~CrA which have an inclination $i$ as large as 60$^\circ$. We  explored various scenarios that could produce such closure phase signals.

\subsubsection{Inclination effects}

We first tried to quantify the closure phase signals induced by a projection effect which is only due to an inclination that is different from face-on. For this purpose, we used radiative transfer models \citep{Klarmann2018tcl..confE..84K} to simulate closure phase signals with different grain populations. We fixed the inclination to 45$^\circ$ and had the grain size vary between 0.1~$\mu$m and 1~mm. At maximum, closure phases of $\pm$ 10$^\circ$ are obtained. Looking at our histogram of closure phase variations (Fig.~\ref{fig:CP}), we can conclude that inclination effects with some grain size distributions (see Sect.~\ref{sect:grainsize}) can explain the closure phase variations below 25$^\circ$ (81\% of our sample). For the highest closure phase signals, however, other scenarios should be invoked.
\begin{figure*}[t]
    \centering
    \includegraphics[width=11cm]{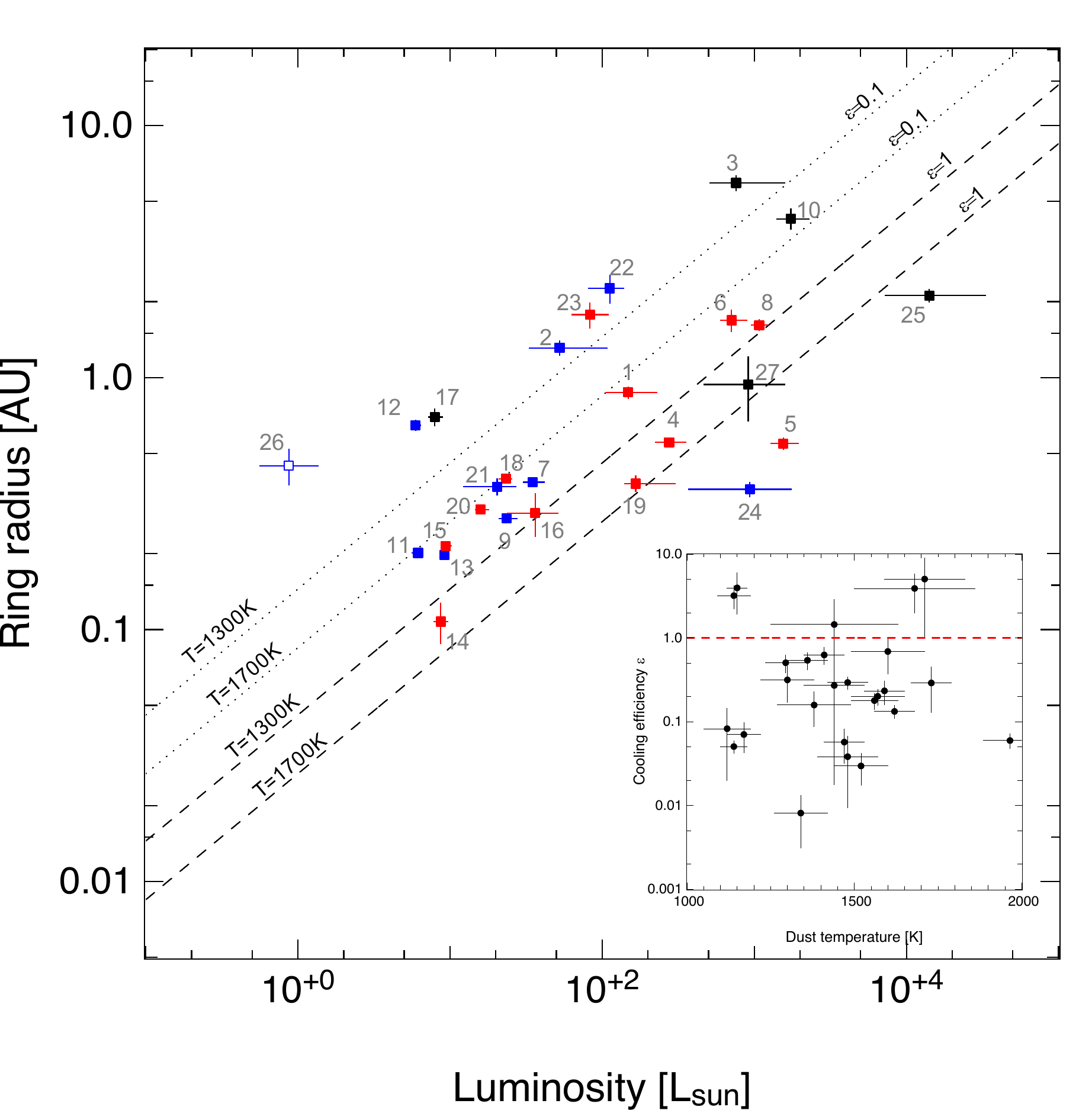}
    \caption{Ring half-flux radius vs. luminosity for group I (blue), group II (red) and unclassified (black) targets. The dashed and dotted lines show the relationship for the labeled temperature and cooling efficiency of the dust $\epsilon_g$ of an optically thin inner cavity model. The inset shows the distribution of values of the dust cooling efficiency $\epsilon$ derived from our ring half-flux radii and dust temperatures. The empty symbol (source\,\#26) corresponds to R\,CrA, a low-quality source in \cite{Vioque2018A&A...620A.128V}. The numbering refers to Table~\ref{tab:starproperties}.}
    \label{fig:RLrelation}
\end{figure*}

\subsubsection{Azimuthal modulation}

In order to mimic the environment's macro-structures as hot spots or arcs, or a binary companion (for 5 among the 6 targets exhibiting a high closure phase variations, \cite{Vioque2018A&A...620A.128V} mark them with a $Bin$ flag), we used the same MCMC approach as previously, and a ring model with an azimuthal modulation described in L17. We took into account the closure phase signals for the fit and the $\chi^2_r$ computation. As expected, the higher the closure phase signals, the stronger the $\chi^2_r$ improvement when considering an azimuthal modulation: this improvement reached a factor 14 for HD~45677, 21 for HD~98922, 6.5 for HD~144668 and 3 for HD~144432. The best-fit models exhibit a brighter, more inclined, and flattened rim as displayed in Appendix D of L17. Two targets (HD~45677 and HD~144668) are best modeled with a skew orientation as expected from a pure inclination effect; these targets have a high inclination of 50-60$^\circ$. Two others (HD~98922 and HD~144432) whose inclinations are  smaller (30-40$^\circ$) are best modeled with a skew angle different from the PA. For HD~179218, the fit does not converge properly, which is not surprising looking at the "flat" visibility curve (Fig.~B.5) and the findings of \citet{Kluska2018ApJ...855...44K}: the latter authors combine optical, near- and mid-infrared high resolution data and find that the near-infrared emission radius is about 30 times larger than the theoretical sublimation radius for a dust temperature of about 1800~K. The authors invoke quantum heated dust particles (e.g. polycyclic aromatic hydrocarbons grains) at large radii to explain this unusual extended emission with such a high temperature. Azimuthal modulation could explain the highest closure phase signals.

\section{Discussion}
\label{sect:disc}

\subsection{Size-luminosity relation}

The size-luminosity diagram explores the correlation between the location of the K-band emission and the strength of the radiation field of the central star. The current picture suggests a strong $R$~$\propto$~$L^{1/2}$ correlation at near-infrared wavelengths, which can be explained by the fact that the emission in this spectral range is dominated by the dust located close to the dust sublimation radius, whose position itself is primarily constrained by the luminosity of the central star \citep{Monnier2005}. This correlation invokes the model of an "optically thin" inner gaseous cavity between the star and the inner radius of the disk. Additionally, to explain the deviation of undersized disks from the correlation law, an alternative model of an "optically thick" inner gaseous cavity is proposed, in which the optically thick gas may shield the dust from direct stellar irradiation, thus bringing the sublimation radius closer to the star.\\

Bringing our GRAVITY observations together with the Gaia DR2 distances (Table~\ref{tab:starproperties}), we increase, by a factor of two with respect to the latter authors, the homogeneous sample of $K$-band disk sizes and significantly improve the coverage of the $\sim$10$^3$\,$L_\odot$ region (Fig.~\ref{fig:RLrelation}). Despite some scatter, we confirm the $R$\,$\propto$\,$L^{1/2}$ correlation around 10$^1$\,$L_\odot$. In the 10$^2$--10$^4$\,$L_\odot$ region (early A to late B spectral type), we observe a larger scatter of the data points around the $L^{1/2}$ slope, which does not clearly favor either the "optically-thick" or the "optically-thin" inner cavity scenario. Such a scatter is in clear agreement with the PIONIER findings (L17).\\


\begin{figure*}[t]
        \centering
        \includegraphics[width=11cm]{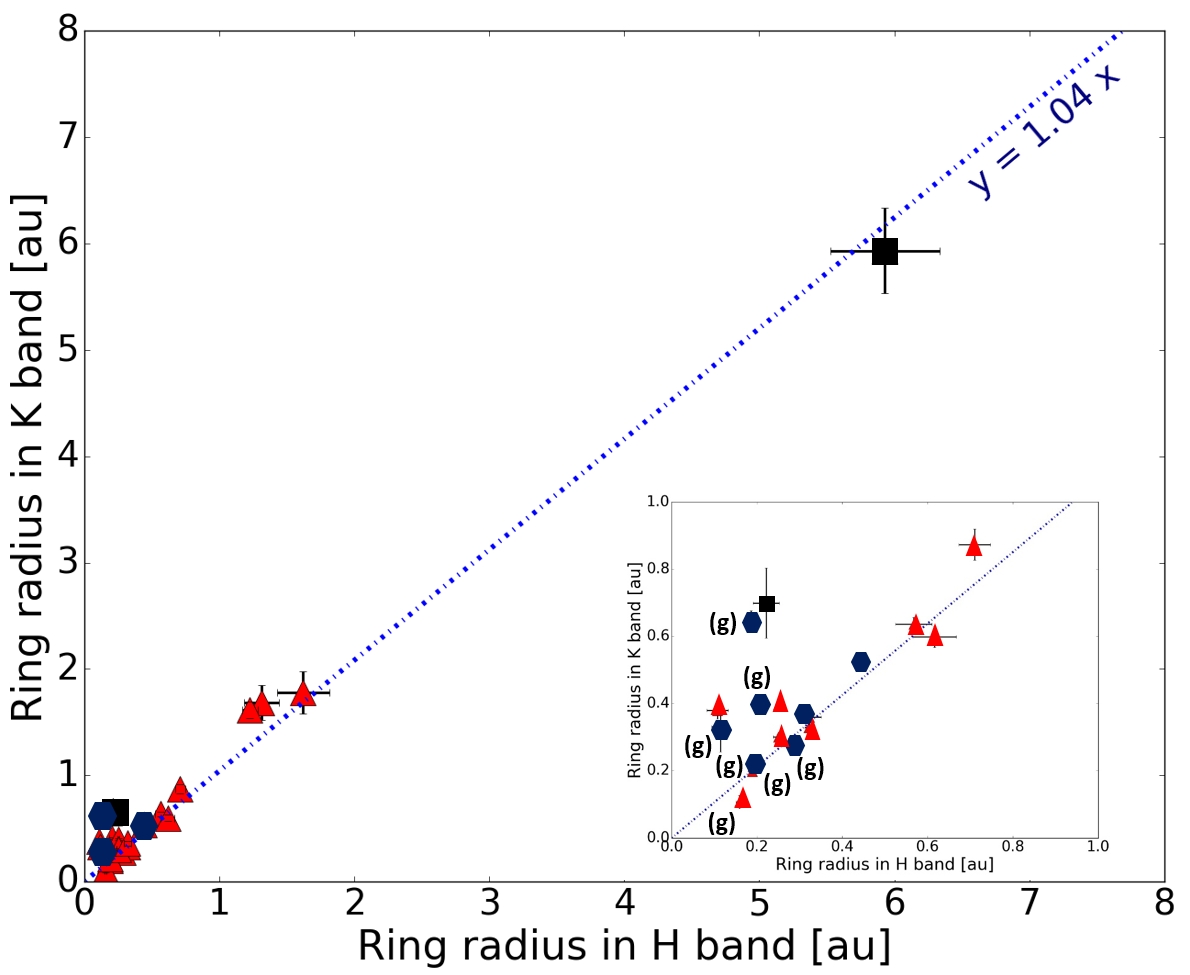}
    \caption{Comparison of half-flux radii of ring models in H and K-bands as determined by  PIONIER and GRAVITY observations (see Table E.1). Blue diamonds denote group-I sources, while the red triangles denote group-II sources. The gapped sources are marked with (g). The dash-dotted line is the resulting linear regression. The inset is a zoom on the smallest ring radii.}
    \label{fig:sizesHK}
\end{figure*}

In the simple model of a passively irradiated disk with an optically thin inner cavity, the dust radius follows \citep{dullemond01}
\begin{equation}
    R(\epsilon) = \left(\frac{L_\star}{16\pi\epsilon\sigma T_{g}^4}\right)^{1/2},
    \label{diskradius}
\end{equation}
where $T_{g}$ is the temperature of the dust grain with a cooling efficiency $\epsilon$, and $R(\epsilon)$ is the stellocentric radius of the K-band emission (which can be different from the sublimation radius). Here, we assumed a backwarming coefficient equal to one \citep{Kama2009}. In Fig.~\ref{fig:RLrelation}, our measurements can be compared to typical trend lines plotted for boundaries of 1300\,K and 1700\,K in temperature as well as 0.1 and 1 in cooling efficiency. Using the same assumption for the disk model, we estimate the dust grain cooling efficiency $\epsilon$ for each object of our sample by taking into account the results of Table~3 (see inset of Fig.~\ref{fig:RLrelation}). The majority of our sources exhibit a dust grain cooling efficiency $\epsilon$ between roughly 0.1 and 1. Although cooling efficiencies larger than one are also possible, in particular for carbon-poor silicate dust and especially forsterite, this is found more at the mid-infrared rather than at near-infrared wavelengths. Hence, in our context, a cooling efficiency larger than one may simply indicate that the assumed disk model is incomplete and wrong.
\vspace{0.15cm} \\

A few sources in our sample merit an individual commentary:
\begin{itemize}
    \item R\,CrA (\#26) belongs to the low-quality sample of \cite{Vioque2018A&A...620A.128V}. The spread in luminosity reported in the literature for this source is larger than the current error bar, which means that this object could move by some extent towards the right part of the diagram.
    \item HD\,85567 (\#5), HD\,259431 (\#24), and PDS\,27 (\#25) appear significantly below the $\epsilon$~=~1 trend line with a rather compact emission size in comparison with their bolometric luminosity. On one side, if we consider their young age of $\sim$0.04--0.4\,Ma (see their position in Fig.~\ref{fig:HR}) and their strong accretion rate of $\sim$10$^{-6}$--10$^{-4}$\,$M_\odot$/yr$^{-1}$ \citep{fairlamb15,Reiter2018}, these sources are likely still in the active disk phase where the accretion luminosity strongly contributes to the energy output close to the star. The inner cavity might be filled with optically thick gas, in part detected with GRAVITY, shielding the dust. With these considerations, the assumption that the characteristic size of the emission is to first order set by the luminosity of the central star might not be valid. The viscous accretion produces an emission that is much closer to the star than in the case of the passive disk with an "optically-thin cavity", which results into the measurement of a more compact circumstellar component.
    \item On the other hand, this must be interpreted in the context of the aforementioned large scatter of the ring radius for high luminosity: indeed HD\,95881 (\#6), HD\,98922 (\#8), and HD\,190073 (\#23) have similar luminosities ($\log L$~$\sim$~3) and masses, and share roughly the same evolutionary track as HD\,259431 (\#24) and HD\,85567 (\#5), but their position in the size-luminosity diagram is quite different, and even opposite to the later ones for  HD\,190073 (\#23) which is above the $\epsilon$~=~0.1 trend line. Although there are only upper limits for the accretion rate found in the literature for HD\,95881 (\#6) and HD\,98922 (\#8) \citep{fairlamb15}, HD\,190073 (\#23) is a strong accretor with $\sim$10$^{-5}$\,$M_\odot$/yr$^{-1}$ yet it is nonetheless found with a large $K$-band characteristic size.
\end{itemize}

\subsection{Can we constrain the grain size distribution of the inner disk?}
\label{sect:grainsize}

Even if there is a strong degeneracy between the inner rim position and the grain populations in terms of composition and size, and if the inner rim position defined by the near-infrared emission can be different from the physical inner rim of the dust disk for a given grain population (Klarmann et al., subm.), we can use our GRAVITY measurements to look for trends:
\begin{itemize}
\item The ring half-flux radii determined by GRAVITY and PIONIER (given in Table E.1 in Appendix E) are comparable at a 3-$\sigma$ level for most of our targets (Fig.~\ref{fig:sizesHK}). The maximum of the K-band emission appears farther away than the H-band one, which is consistent with a K-band emission cooler than the H-band one (see Sect. 5.4). The outliers observed at the shortest radii generally correspond to gapped sources for which the halo contributions are expected to be larger than for ungapped sources, and these can be different in the H-band than in the K-band. As mentioned in Sect. 5.2, the halo contribution can only be probed accurately with the very short baselines that are generally not covered in our (u, v) planes.
\item The best models in the H and K-bands are wide and smooth rings, and the closure phase distribution (Fig.~\ref{fig:CP}) displays a median value larger than 11$^\circ$, which favors wedge-shaped inner rims and disfavors mono-size grain distributions since the latter produce very thin, sharp rings as a result of all grains being sublimated at the same stellar distance. As reported by Klarmann et al. (submitted), wedge-shaped inner rims can be reproduced with grain distributions of two different kinds or more, that is, with grains with a cooling efficiency close to 1 that survive near the inner rim and grains with a smaller cooling efficiency and a temperature close to the sublimation temperature that survive further away. Such distributions, meaning those with sub-micron or micron sized grains, produce maximal closure phases of $\pm$~10$^\circ$ and a halo contribution at a level of 10-20\% due to a high level of scattered light. Adding larger grains or considering a MRN grain size distribution \citep{Mathis1977ApJ...217..425M} leads to a decrease in the closure phases.
\item The cooling efficiency depends on the chemical composition of the dust grain \citep{Kama2009}, as well as on the grain size and radiative equilibrium temperature. Grains significantly smaller than the observing wavelength become poor emitters\footnote{With $\epsilon_g$\,$\propto$\,$2\pi r/\lambda$ where $r$ is the grain radius \citep{Chiang1997}}, thus leading to a reduced cooling efficiency. According to the location of our sources in the size-luminosity diagram (Fig.~\ref{fig:RLrelation}), the inner disk K-band emission is dominated by grains with low cooling efficiency, although the origin of this low value -- the intrinsic opacity of the grain or sub-micron size -- cannot be definitely disentangled here. This scenario does not exclude, of course, the presence of a large reservoir of grains with cooling efficiency close to unity -- due, for instance, to micron-sized (or larger) grains -- but this population would not be sensed by GRAVITY if they already had become sedimented towards the optically thick mid-plane.
\end{itemize}

\subsection{Can we probe the inner disk morphology?}
\label{groups}

In the Meeus classification \citep{Meeus2001}, disks of Herbig Ae/Be stars are classified as group I {\it } for flared disks and group II for flat
disks. \cite{Maaskant2013A&A...555A..64M} used spatially-resolved mid-infrared imaging to suggest that group I disks are preferentially gapped structures, with the resulting large puffed-up wall at the inner edge of the outer disk making the comparative size of the mid-infrared emission larger in relation to the flat, preferentially continuous, disks of group II.
Using mid-infrared interferometry, \cite{Menu2015A&A...581A.107M} find that the presence of a gap is not a unique feature of group I disks but may also be present in group II disks, albeit with a smaller width.

Looking at Fig.~\ref{fig:RLrelation} from the perspective of a group-I/group-II dichotomy may graphically suggest that, for a given luminosity class, the group-I (blue symbols) characteristic sizes generally appear larger than those in group-II (red symbols), particularly in the 10--100\,$L_\odot$ range. However, this interesting hypothesis, which would have important implications in terms of disk structuring, is difficult to confirm beyond any reasonable doubt. Future work on the characteristic of disk size and infrared classification of these objects may shed more light on this area.

From the 27 objects in our GRAVITY sample, 19 of them have a mid-infrared disk size counterpart measured with MIDI. Since the mid-infrared sizes reported by \cite{Menu2015A&A...581A.107M} are the result of modelling face-on disks and neglect inclination effects, we compute an {\it a posteriori} correction factor that is to be applied: first, we check their reported (u, v) spatial frequency planes to verify if a specific position angle has been preferentially probed by the MIDI baseline. When the (u, v) coverage is qualitatively uniform, we propose to correct the published mid-infrared size by the multiplicative factor $f$~=~2~-~$\cos$~$i$, with $\cos$\,$i$ measured with GRAVITY. For two particular sources, HD\,95881 and HD\,142666, the limited MIDI (u, v) coverages indicate a preferential position angle along the ellipse minor axis, we apply a multiplicative factor of $f$~=1/$\cos$~$i$ (Table~\ref{tab:group12}). 

\begin{table}[h]
\centering
\caption{K-band and N-band characteristic sizes. Near-IR sizes correspond to half-flux diameter ring model. Mid-infrared data are the half-flux diameter adapted from \cite{Menu2015A&A...581A.107M}. Multiplicative factor $f_c$ is applied to the measured N band characteristic diameter.}
{\tiny
\begin{tabular}{c l l l c c }
\hline \hline
\# & Object & FWHM\,(K) & FWHM\,(N) & $\cos$\,$i$\,($f_c$) & Group \\
&       & [mas]     & [mas]     &   \\ \hline
1& HD\,45677    & 19.1$\pm$0.4  & 77    $\pm$ 2   & 0.49$\pm$0.01\,(1.51)  & -- \\
2& HD\,95881    & 2.89$\pm$0.2  & $<$11.1             & 0.61$\pm$0.03 & II \\
3& HD\,98922    & 4.69$\pm$0.11 & 21.7  $\pm$ 0.4 & 0.66$\pm$0.01\,(1.34) & II\\
4& HD\,144432   & 2.76$\pm$0.07 & 27  $\pm$ 1     & 0.68$\pm$0.02\,(1.21) &II\\
5& HD\,144668   & 4.08$\pm$0.1  & 13.1  $\pm$ 0.1 & 0.56$\pm$0.01\,(1.44) & II\\
6& HD\,150193   & 5.64$\pm$0.13 & 42.0  $\pm$ 0.8 & 0.68$\pm$0.03\,(1.32) & II\\
7& HD\,158643   & 5.26$\pm$0.12 & 34.5  $\pm$ 0.6 & 0.46$\pm$0.01\,(1.54) & II\\
8& HD\,163296   & 5.90$\pm$0.14 & 20  $\pm$ 1     & 0.76$\pm$0.01\,(1.24) & II\\
9& HD\,142666   & 1.45$\pm$0.26 & 66.9  $\pm$ 3.6 & 0.55$\pm$0.04 & II\\
10& HD\,259431  & 1.00$\pm$0.03 & 9.2   $\pm$ 0.2 & 0.88$\pm$0.02\,(1.12) & I\\
11& HD\,139614  & 9.57$\pm$0.43 & 49.3  $\pm$ 0.8 & 0.71$\pm$0.04\,(1.29) & I\\
12& HD\,38120   & 6.47$\pm$0.15 & 91  $\pm$ 4     & 0.66$\pm$0.02\,(1.34) & I\\
13& R\,CrA      & 9.41$\pm$0.75 & 41.7  $\pm$ 0.6 & 0.52$\pm$0.04\,(1.48) & I\\
14& HD\,100546  & 5.02$\pm$0.12 & 521   $\pm$ 18  & 0.65$\pm$0.01\,(1.35) & I\\
15& HD\,142527  & 2.52$\pm$0.06 & 19.1  $\pm$ 0.5 & 0.90$\pm$0.01\,(1.1) & I\\
16& HD\,169142  & 6.47$\pm$0.46 & 227   $\pm$ 40  & 0.8$\pm$0.2\,(1.2) & I\\
17& HD\,179218  & 17.0$\pm$2.10 & 78  $\pm$ 4     & 0.59$\pm$0.13\,(1.41) & I\\
18& HD\,97048   & 4.18$\pm$0.10 & 68    $\pm$ 3   & 0.72$\pm$0.02\,(1.28) & I\\
19& HD\,135344  & 2.96$\pm$0.07 & 8.1   $\pm$ 0.4 & 0.73$\pm$0.02\,(1.27) & I\\
\hline
\end{tabular}\label{tab:group12}
}
\end{table}

\begin{figure}[h]
    \centering
    \includegraphics[width=\columnwidth]{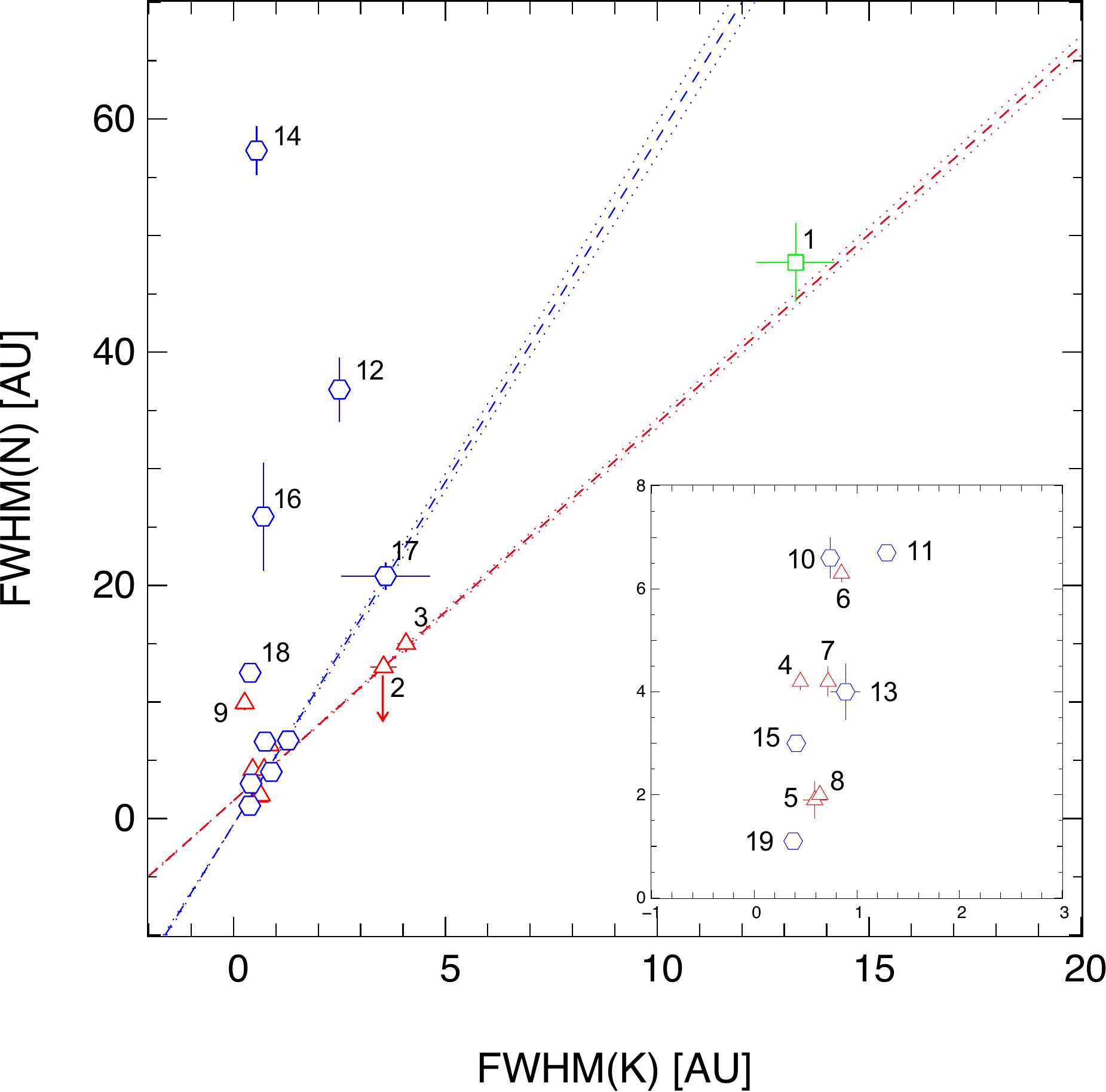}
   \caption{Size-size diagram of  FWHM in N- and K-bands for group I (blue), group II (red), and unclassified (black) sources. The numbering refers to Table~\ref{tab:group12}. Blue and red dashed lines correspond to an error-bar-weighted linear regression on the group I and group II sources, respectively: $y$~=~$ax$~+~$b$ with $a_I$=5.87$\pm$0.14, $b_I$=-0.50$\pm$0.08, $a_{II}$=3.23$\pm$0.04, $b_{II}$=1.54$\pm$0.09. The dotted lines delimit the uncertainty on the slope parameters. The inset shows a zoom of the overlap region (see text for details).
}
    \label{fig:KvsN}
\end{figure}

\begin{figure*}[t]
        \centering
        \includegraphics[width=1\columnwidth]{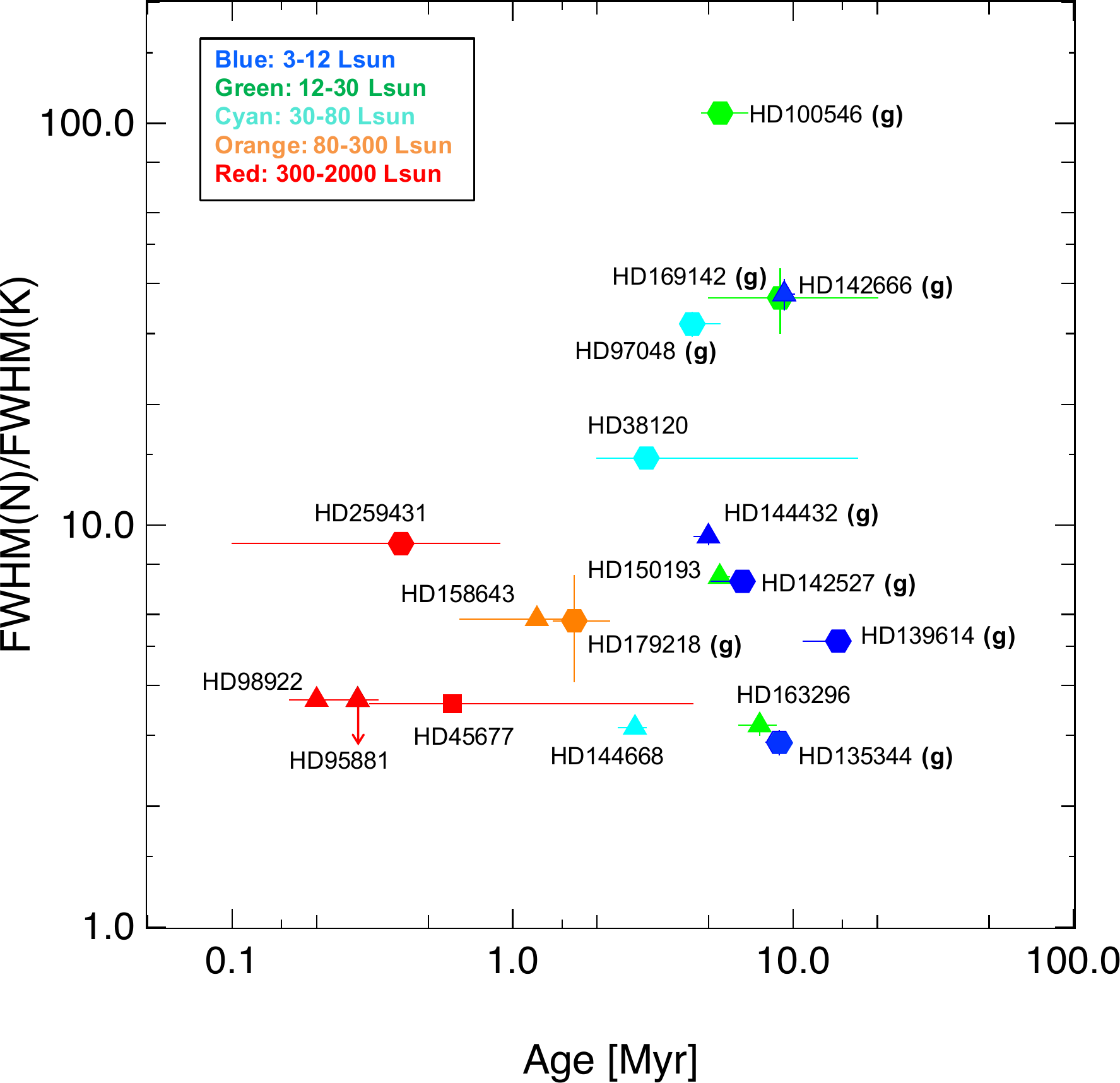}
        \includegraphics[width=1\columnwidth]{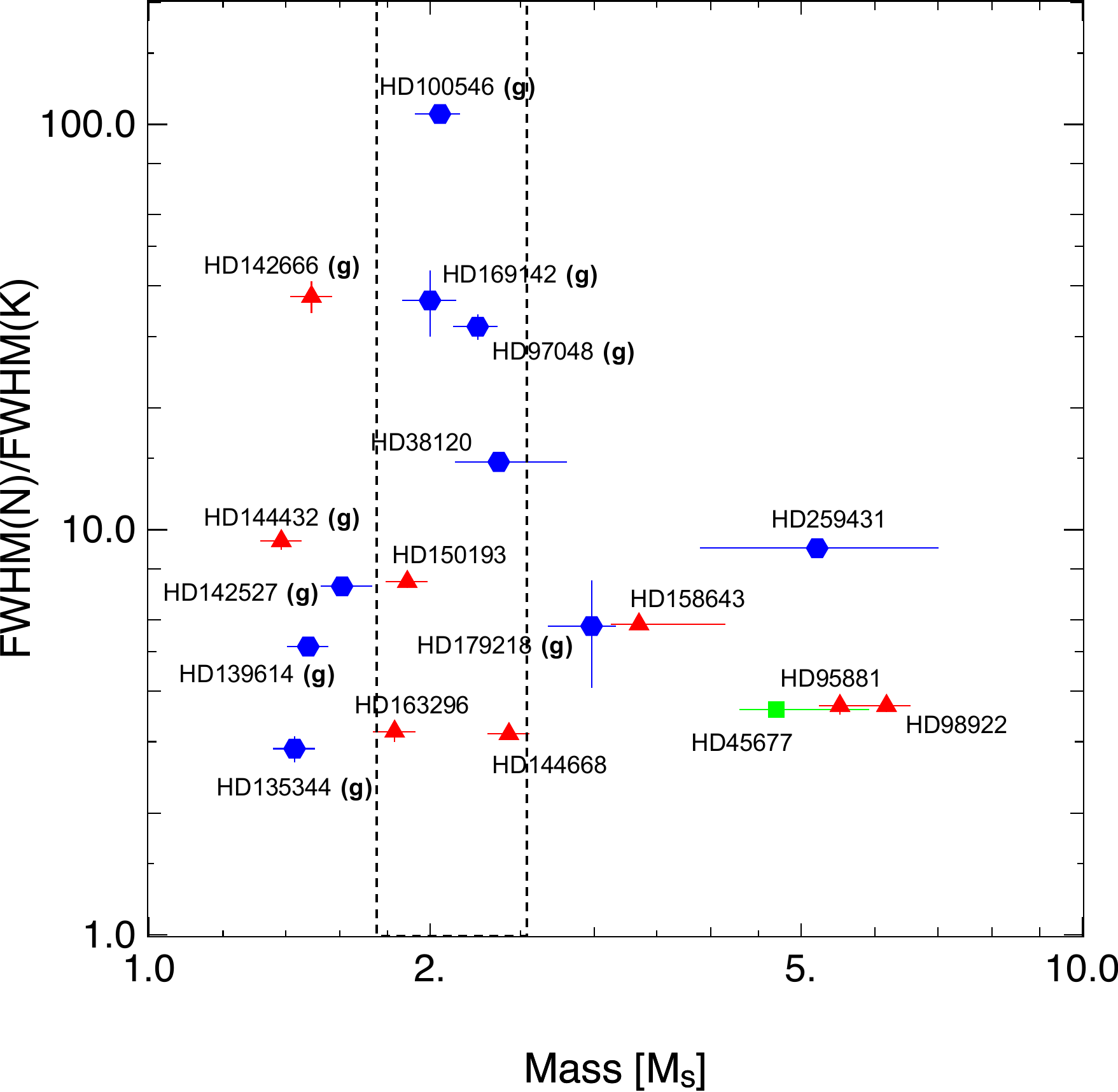}
    \caption{N-to-K ratio as function of age and mass.
     The triangle symbol shows group-II sources, diamond symbol shows group-I sources and the square symbol shows objects with no classification (i.e., HD\,45677).
    In the left plot, the different colors correspond to different luminosity classes as shown in the inset. In the right plot, blue diamonds identify group-I sources and red triangles group-II sources. Identified gapped sources by high angular resolution observations are marked with (g). The object R\,CrA (\#13 in Table~\ref{tab:group12}) is not represented here as no age estimate is reported in \cite{Vioque2018A&A...620A.128V}.}
    \label{fig:size-age}
\end{figure*}

In the size-size diagram of Fig.~\ref{fig:KvsN}, we compare the K-band and N-band sizes for sources classified as group I or group II according to \cite{Meeus2001,VanBoekel2005,Juhasz2010}. In the first order, the characteristic sizes of the K-band and N-band  increase proportionally. At the small-scale level (i.e., $<$2\,au in K-band and $<$8\,au in N-band) there is no strong segregation between the groups I and II, and an overlap is observed, as found by \citet{Maaskant2013A&A...555A..64M}. However, at a larger scale, the group-I sources appear more clearly on the upper part of the diagram with respect to the group-II sources, as demonstrated by the plotted linear regression laws for the two populations.

\subsection{Evolution of the inner disk structure}
\label{evol}

An interesting question underlying to the group I/group II disk classification is the possible evolutionary link between these two different populations. \cite{Maaskant2013A&A...555A..64M} advance the hypothesis that group I (presumably gapped) and II (presumably continuous) disks may not be connected via a common evolutionary path, but represent, rather, two spatial configurations for independent disks present in different objects at a similar age. \cite{Menu2015A&A...581A.107M} also propose an alternative evolutionary scenario for group I and II disks: either  each group might follow a distinct evolutionary path from ungapped to gapped disk, or flat gapped disks could later evolve into flared disks with larger gaps. Recent ALMA results provide clear evidence that gaps in disks -- whatever the origin of their formation -- are not only a feature specific to the inner disk region but one that can also be found in a concentric arrangement out to large distances \citep{Zhang_2018}. Since the one-gap scenario might need to be revisited, the same may occur with the original group I/II classification.

We use the absolute ages and masses provided by \cite{Vioque2018A&A...620A.128V} (Table 1) to study the relationship between the age of the system and the ratio of the measured FWHM values in the N and K-bands (N-to-K ratio).

\begin{itemize}
\item From Fig.~\ref{fig:size-age}-left, we see that all the sources with a N-to-K ratio larger than $\sim$10 are classified as group-I (in the sense of Meeus 2001), except for HD\,142666. However, this last source is found by \cite{Schegerer2009} to have a gap of $\sim$0.5\,au, which makes its position in our diagram relatively coherent. The N-to-K ratio could be used in part as a proxy to assess the group-I classification.
\item From the plots of Fig.~\ref{fig:size-age}, a marked distinction in the N-to-K ratio appears when looking at the luminosity class, and, correspondingly, that of mass. The brightest, and most massive objects (orange and red colors) show a ratio less than $\sim$10, whereas objects with lower luminosities (blue to cyan) clearly have a stronger scatter in the N-to-K ratio with values reaching above $\sim$100.
\item With no further consideration of the spectral type, we observe for our sample an increasing scatter in the N-to-K ratio for sources older than $\sim$1\,Ma. 
\end{itemize}

The distribution of group I and group II sources does not clearly point toward a segregation between the two populations based on their "absolute" age or mass, with both types of objects found among ranges in question. However, there is a mass bias in the age determination of pre-main sequence stars through stellar interior modeling. As a consequence, in order to correctly interpret Fig.~8, we should focus more on specific iso-mass paths (see Fig. 1) along which the "relative" ages are more relevant. Looking at the targets at about 2~M$_\odot$ (see the dotted box in Figs. 1 and 8-right), we see that the more-evolved sources (in this case group I) closer to the Zero-Age Main-Sequence (ZAMS; Fig. 1) exhibit a larger N-to-K ratio (Fig. 8-right), whereas the less-evolved sources (in this case group II) show a smaller N-to-K ratio. Interestingly, for the objects of about 1.5 M$_\odot$ in Fig.~8-right, we also find that (with the exception of HD~139614), the more-evolved sources toward the ZAMS (in this case group II) display a larger N-to-K ratio than the less-evolved objects (in this case group I). Note that for 1.5 M$_\odot$ all our sources are reported as gapped disks. The limited size of our sample makes it difficult to propose a global evolutionary picture for now, but such a plausible and important trend which would imply an increase of the N-to-K ratio with age should be confirmed through the undertaking of additional measurements of K- and N-band sizes aimed at populating the plots of Fig.~8 with, for instance, all the sources of the iso-mass 5~M$_\odot$ in Fig.1.

\subsection{Gap formation scenarios}

Photo-evaporation \citep{Armitage2011} is one of the key processes of disk dispersal via the formation of gaps and disk inner holes. Heating by the central star through radiations in the Extreme Ultra-Violet (EUV), Far Ultra-Violet (FUV), and in the X-rays induces the splitting of the disk and the formation of a gap with a peak efficiency at the critical radius $r$$_c$$\sim$\,0.1--0.2\,$r_g$ \citep{Gorti2009}, where $r_g$~=~$GM_{\ast}/c_s^2$ is the radius where the heated gas in the disk becomes gravitationally unbound with regard to the central star. In the EUV/FUV/X-ray photo-evaporation scenario, the formation of gap takes $\sim$4~Ma (see Fig. 8 in \cite{Gorti2009}) for solar-type masses, and is likely shorter for masses larger than 3~$M_\odot$ (see Fig. 12 of \cite{Gorti2009}; \cite{Ercolano2017}). Once the gap is formed, the inner disk is rapidly depleted in a very short viscous timescale of $\sim$10$^5$ years \citep{Gorti2009}. Over the course of this interval, the result is the formation of the transitional disk with a void inner cavity.

Using our GRAVITY observations, we can directly compare the characteristic size of the K-band emission to the theoretical critical radius \citep{Gorti2009}
\begin{equation}
    r_c \simeq 1.05\left(\frac{M_\ast}{M_\odot}\right)\left(\frac{T}{10^4\,K}\right)^{-1}\,\rm au \label{eq-photoevap}
\end{equation}
where $T$ is the temperature of the EUV/FUV/X-rays heated gas. 

\begin{figure}[h]
    \centering
    \includegraphics[width=0.95\columnwidth]{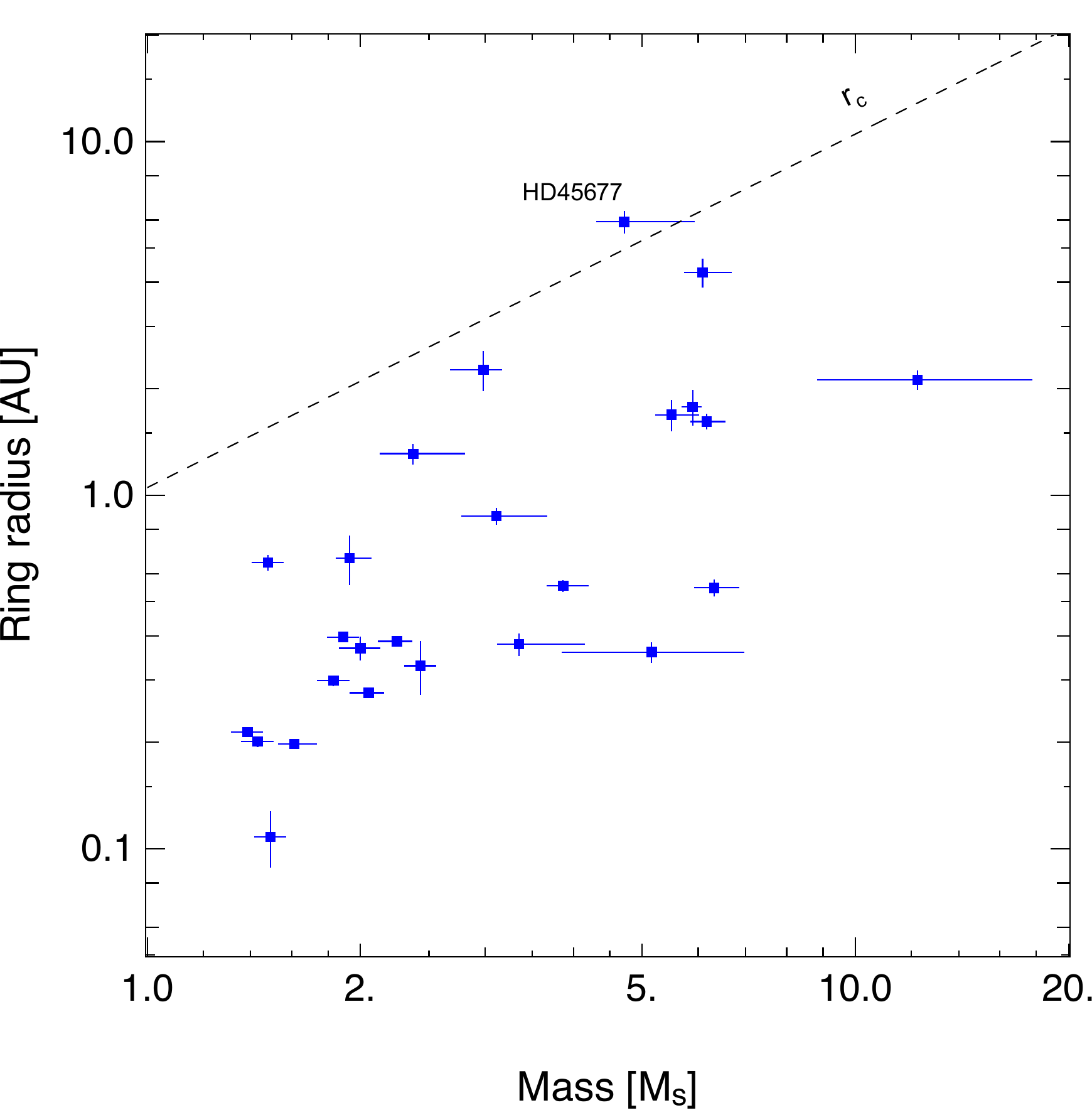}
    \caption{Half-flux ring radius as function of mass of central star. Dashed line corresponds to the critical radius $r_c$ given by Eq.~\ref{eq-photoevap} as T~=~10$^4$~K.}
    \label{fig:photoevap}
\end{figure}

With the exception of HD\,45677, all our sources lie below the critical radius for a standard $T$\,$\sim$10$^4$\,K temperature (Fig~\ref{fig:photoevap}). When looking at our gapped sources in Fig.~\ref{fig:size-age} with an age of $\sim$5--10\,Ma, and assuming the aforementioned photo-evaporation scenario, we could have expected these objects to already have been in the transitional stage with the inner rim of the disk beyond the critical radius, but this is not the case according to Fig.~\ref{fig:photoevap}. According to this model, EUV/FUV/X-ray photo-evaporation would have had a negligible or no effect for these disks. Alternatively, one may surmise that the timescales for photo-evaporation are significantly longer as found for EUV photo-evaporation, or that the process of gap formation in these discs is indeed dominated by the dynamical clearing of young planets. For the objects with an age of $\sim$1\,Ma and younger, the smaller sizes compared to $r_c$ are compatible with a scenario where photo-evaporation had just started. Finally, given the approximations in Eq.~\ref{eq-photoevap}, the position of HD\,45677 with respect to the $r_c$ slope remains uncertain. However, when combining the different graphic results of Sect.~\ref{sect:disc}, this source appears consistent with the evolutionary stage of a young, flat and ungapped disk. This could be corroborated by photometry and SED analysis following \citet{Meeus2001} and \citet{VanBoekel2005}.

\subsection{Inner and outer disk (mis)alignment}

Connecting the inner and outer parts of  protoplanetary disks are essential for understanding all the dynamic effects involved, as well as the implications for planet formation in particular. While the recent images of scattered light produced by SPHERE and images from ALMA in the millimetric range have allowed us to determine orientations for the outer disks accurately, near-infrared interferometry is a relevant means to measure these inclinations and these position angles of the inner disks. 

We aim to compare our inner disk orientations with the outer disk orientations derived from imaging using either SPHERE or ALMA. Since the ALMA and SPHERE images have a variety of spatial resolutions, in order to simplify analysis and avoid the over-interpretation of small variations between inner and outer inclinations and position angles, we determine a uniform uncertainty on these values that is to be applied to all objects. We generate a family of synthetic observations of thin disks with known inclination and position angles for the nearly edge-on ($\sim$80$^\circ$) and nearly face-on ($\sim$10$^\circ$) mimicking the lowest spatial resolution data in our target sample [i.e., ALMA data of HD~97048 with $\sim$0.5 $^{\prime\prime}$ resolution, \citep{vanderPlas17}]. For both the nearly face-on case (5$^\circ$ to 15$^\circ$) and the edge-on one (75$^\circ$ to 85$^\circ$), we vary the inclination in steps of 1$^\circ$. After convolving the model thin disks with the $\sim$0.5$^{\prime\prime}$ beam and producing the synthetic observations, we fit Gaussians to the resulting images to recover the major and minor axes and calculate the inclination and position angles. The inclination angles of both the edge-on and face-on cases can be reliably recovered to a level within 4$^\circ$. While the position angles of the edge-on case can be recovered to within 2$^\circ$, the face-on case carried more uncertainty, with only a 5$^\circ$ accuracy.  Since most of our sources are closer to face-on than edge-on, we  adopt a uniform and conservative 5$^\circ$ uncertainty for the inclination and position angles of all sources. Since the SPHERE and ALMA orientations are consistent with each other amid our conservative error bars of 5$^\circ$, we consider the average between the two determinations. 

\begin{figure}[h]
        \centering
        \includegraphics[width=9.5cm]{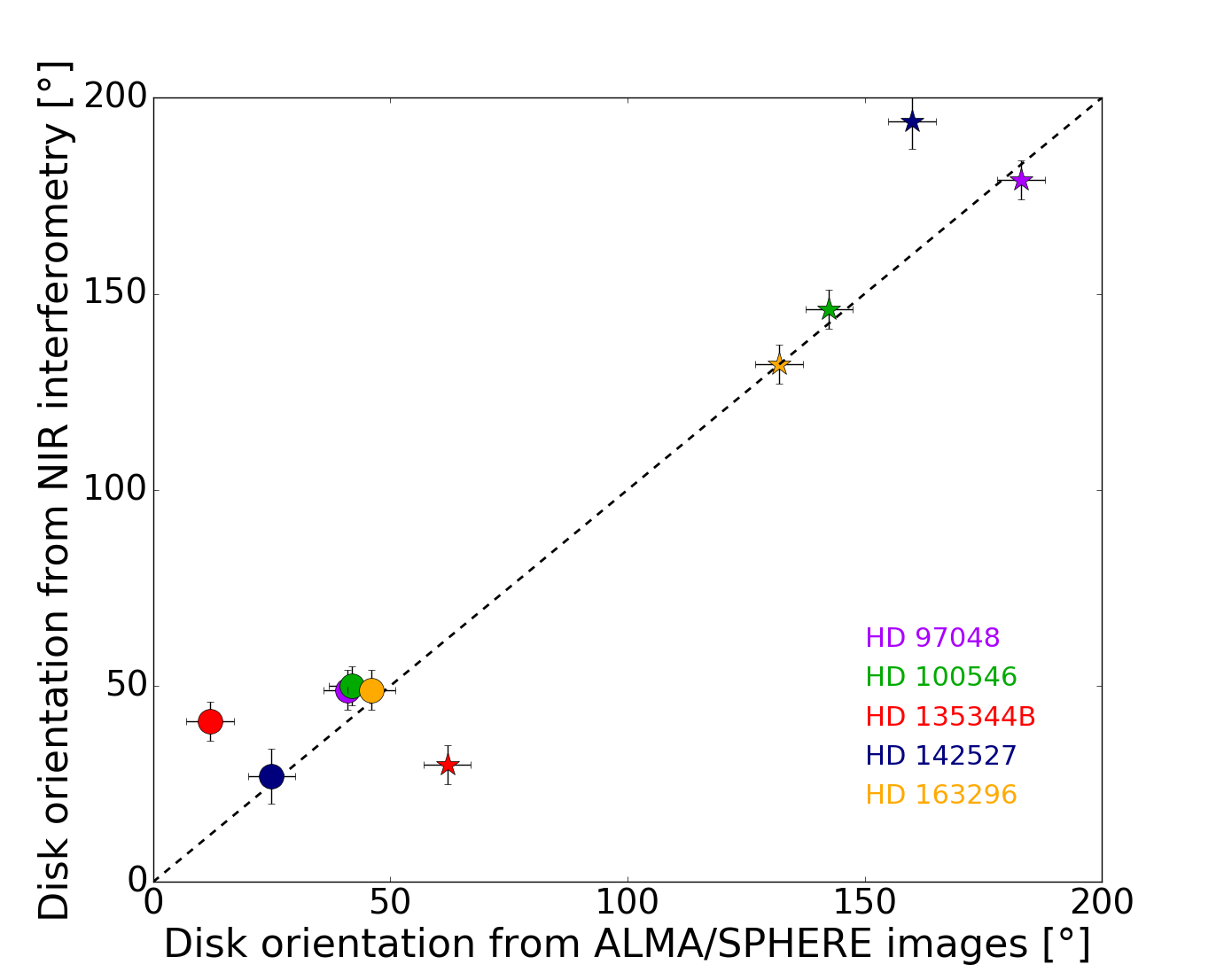}
    \caption{Inclination (circles) and position angle (star) of inner disk derived from our GRAVITY interferometric measurements as a function of inclination and position angle of outer disk derived from imaging with ALMA and SPHERE$^{(*)}$. Dashed lines show 1:1 relations.\protect\\
    {\footnotesize $^{(*)}$ HD~97048: \cite{vanderPlas17}, \cite{Ginski}; HD~100546: \cite{Pineda},\cite{Sissa}; HD~135344B: \cite{cazzoletti}, \cite{Maire}; HD~142527: \cite{kataoka}, \cite{Avenhaus2018ApJ...863...44A}; HD~163296: \cite{huang}, \cite{Muro}}}
    \label{fig:in-out}
\end{figure}

We can compare these orientations for five targets (the orientations for HD~169142 using GRAVITY are too poorly constrained). We note differences  for HD~135344B and HD~142527 (Fig.~\ref{fig:in-out}). Interestingly, recent images in polarized light of the latter star have revealed features like spirals and dark arcs \citep{Avenhaus2014ApJ...781...87A}. A scenario that has been invoked to explain such features is attributed to shadows cast by an inner disk that is misaligned with respect to the outer disk \citep{Marino2015ApJ...798L..44M}. The authors cited use radiative transfer models to reproduce the NACO observations and they derive a relative inclination between the inner and outer disks of 70~$\pm$~5~$^\circ$ and a position angle of the inner disk of -8~$\pm$~5~$^\circ$. While we determine a small position angle ($\sim$~10~$^\circ$), our near-infrared interferometric observations are not consistent with such a high difference in inclination. It is out of the scope of this paper to address the open questions on the dust structures of this transitional disk \citep{Avenhaus2018ApJ...863...44A}. Increasing the sample for comparing the orientations of the inner and outer disks, especially with regard to targets for which arcs, warps, and shadows have been observed and or for which planets have been detected, could be a key to a better understanding of the interplay between the different components of these complex and dynamical environments, and of the origins of the variety of features seen by disk imaging as well.

\section{Summary and perspectives}

Near-infrared long baseline interferometry is a powerful means to constrain the continuum emission due to the effects of reprocessed light by the dust grains in the circumstellar environments of young stellar objects. We extend the PIONIER study of Herbig Ae/Be stars to the K-band through a resolution with the GRAVITY instrument of the inner disks around 27 targets of a sample that spans a large range of luminosity, mass, and age. The environments of these stars in the near-infrared continuum appear as wide and smooth rings, exhibiting an asymmetry since the measured closure phases are most often non-zero. To probe the disk morphology, a comparison of the K-band FWHM with N-band ones could be used as a proxy to disentangle flat and flared disks since a transition around the N-to-K ratio of about 10 is observed. The brightest, most massive stars show a ratio less than 10, whereas objects with lower luminosities clearly have a stronger dispersion in the N-to-K ratio with values up to 100. As far as disk evolution is concerned, no universal mechanism is clearly in evidence since the different types of disks (flat or flared, with or without gaps) are found ranging from 0.2~Ma to 15~Ma. In looking at the inner positions derived from our GRAVITY measurements, it appears that gap formation processes in the disks are likely related to the dynamical clearing of young planets rather than FUV/EUV/X-ray photo-evaporation. Near-infrared long baseline interferometry can also be used to derive the inner disk orientations that are to be compared with those of the outer disk so as to better understand the origins of the variety of features seen through disk imaging in these complex and dynamical environments.

As perspectives, investigating macrostructures that have proven more complex physically using radiative transfer codes or the presence of undetected close companions to explain the largest closure phases would be a natural follow-up for the current study. Combining GRAVITY and MATISSE observations will be of utmost interest in probing dust distribution and disk mineralogy and understanding the mechanism for disk evolution and the gap formation processes by populating the N-to-K size diagrams with stars of a few solar masses.

\begin{acknowledgements}
We thank the anonymous referee for her/his fruitful comments. GRAVITY is developed in a collaboration by the Max Planck Institute for Extraterrestrial Physics, LESIA of Paris Observatory and IPAG of Universit\'e Grenoble Alpes / CNRS, the Max Planck Institute for Astronomy, the University of Cologne, the Centro Multidisciplinar de Astrofisica Lisbon and Porto, and the European Southern Observatory. We acknowledge the funding of the French National Program of Stellar Physics (PNPS). This work has been supported by a grant from LabEx OSUG@2020 (Investissements d'avenir -- ANR10LABX56).
This research was partially supported by Funda\c{c}\~ao para a Ci\^{e}ncia e a Tecnologia, with grants reference UID/FIS/00099/2013, SFRH/BSAB/142940/2018 (P.G.). R.G.L.  has received funding from the European Union's Horizon 2020 research and innovation programme under the Marie Sklodowska-Curie Grant Agreement No. 706320 and the Grant from Science Foundation Ireland under Grant number 18/SIRG/5597.
\end{acknowledgements}

%
\bibliographystyle{aa} 
\bibliography{reference} 

%

\begin{appendix}

\section{GRAVITY observations}

The log of the observations is given in Table~A.1.

\begin{table*}[h]
\label{tab:obs}
\caption{Observation log of VLTI/GRAVITY observations. $N$ denotes the number of 5-minute long files that have been recorded on the target.}
\centering
\vspace{0.1cm}
\begin{tabular}{c c c c c}
\hline \hline
HD & Date & Configuration & N & Calibrator \\ 
\hline 
37806 & 2019-03-19 & D0-G2-J3-K0 & 5 & HD~64215\\
\hline
38120 & 2019-03-18 & D0-G2-J3-K0 & 8 & HD~37356, HD~38225\\
\hline
45677 & 2016-12-15 & A0-B2-C1-D0 & 4 & HD~45420\\
 & 2016-12-16 & A0-B2-C1-D0 & 5 & HD~45420\\
 \hline
58647 & 2017-03-18 & A0-G1-J2-K0 & 8 & HD~65810\\
& 2017-03-20 & A0-G1-J2-K0 & 9 & HD~65810\\
\hline
85567 & 2017-02-19 & A0-G1-J2-K0 & 4 & HD~76538\\
& 2017-03-17 & A0-G1-J2-J3 & 10 & HD~79447\\
\hline
95881 & 2018-03-02 &  D0-G2-J3-K0 & 5 & HD~90452\\
\hline
97048 & 2017-03-19 & A0-G1-J2-K0 & 6 & HD~82554\\
& 2017-03-20 & A0-G1-J2-K0 & 6 & HD~118934\\
\hline
98922 & 2017-02-21 & A0-G1-J2-K0 & 14 & HD~103125\\
& 2017-03-18 & A0-G1-J2-K0 & 5 & HD~100825\\
\hline
100546 & 2018-04-27 & UT1-2-3-4 & 9 & HD~99264\\
\hline
114981 & 2019-03-19 & D0-G2-J3-K0 & 6 & HD~113776 \\
\hline
135344B & 2018-03-05 & A0-G1-J2-J3 & 8 & HD~132763\\
\hline
139614 & 2019-03-19 & D0-G2-J3-K0 & 3 & HD~148974\\
\hline
142527 & 2017-03-18 & A0-G1-J2-K0 & 5 & HD~143118\\
\hline
142666 & 2018-06-16 & D0-G2-J3-K0 & 9 & HD~148605\\
\hline
144432 & 2018-03-05 & A0-G1-J2-J3 & 5 & HD~132763\\
\hline
144668 & 2017-03-19 & A0-G1-J2-K0 & 7 & HD~143118 \\
& 2017-05-29 & B2-D0-J3-K0 & 9 & HD~143118 \\
\hline 
145718 & 2019-03-18 &D0-G2-J3-K0 & 8 & HD~145809\\
\hline
150193 & 2018-06-15 & D0-G2-J3-K0 & 7 & HD~148605 \\
& 2018-07-07 & D0-G2-J3-K0 & 3 & HD~181240\\
& 2018-08-12 & A0-G1-J2-K0 & 2 & HD~151635\\
& 2018-08-13 & A0-G1-J2-K0 & 6 & HD~150370\\
\hline
158643 & 2017-05-29 & B2-D0-J3-K0 & 6 & HD~163955\\
& 2017-05-30 & B2-D0-J3-K0 & 14 & HD~163955\\
& 2017-08-15 & A0-G1-J2-K0 & 7 & HD~163955\\
\hline
163296 & 2018-07-07 & D0-K0-G2-J3 & 12 & HD~160915\\
\hline
169142 & 2017-08-16 & A0-G1-J2-K0 & 4 & HD~169830\\
\hline
179218 & 2018-07-07 & D0-G2-J3-K0 & 9 & HD~181240\\
\hline
190073 & 2018-06-15 & D0-G2-J3-K0 & 8 & HD~183936 \\
\hline
259431 & 2017-02-20 & A0-G1-J2-K0 & 2 & HD~48784\\
& 2018-03-05 & A0-G1-J2-J3 & 12 & HD~43386, HD~50277\\
\hline
PDS 27 & 2019-03-18 & D0-G2-J3-K0 & 8 & HD~73434\\
\hline
R CrA & 2018-07-07 & D0-G2-J3-K0 & 3 & HD~181240\\
\hline
V1818 Ori & 2019-03-19 & D0-G2-J3-K0 & 6 & HD~34045\\
\hline
\end{tabular}
\end{table*}


\section{GRAVITY data}

The visibilities squared and closure phases observed by GRAVITY are displayed as a function of the interferometric baselines (Fig.~B.1-6).

\begin{figure*}[h]
        \centering
        \includegraphics[width=15cm]{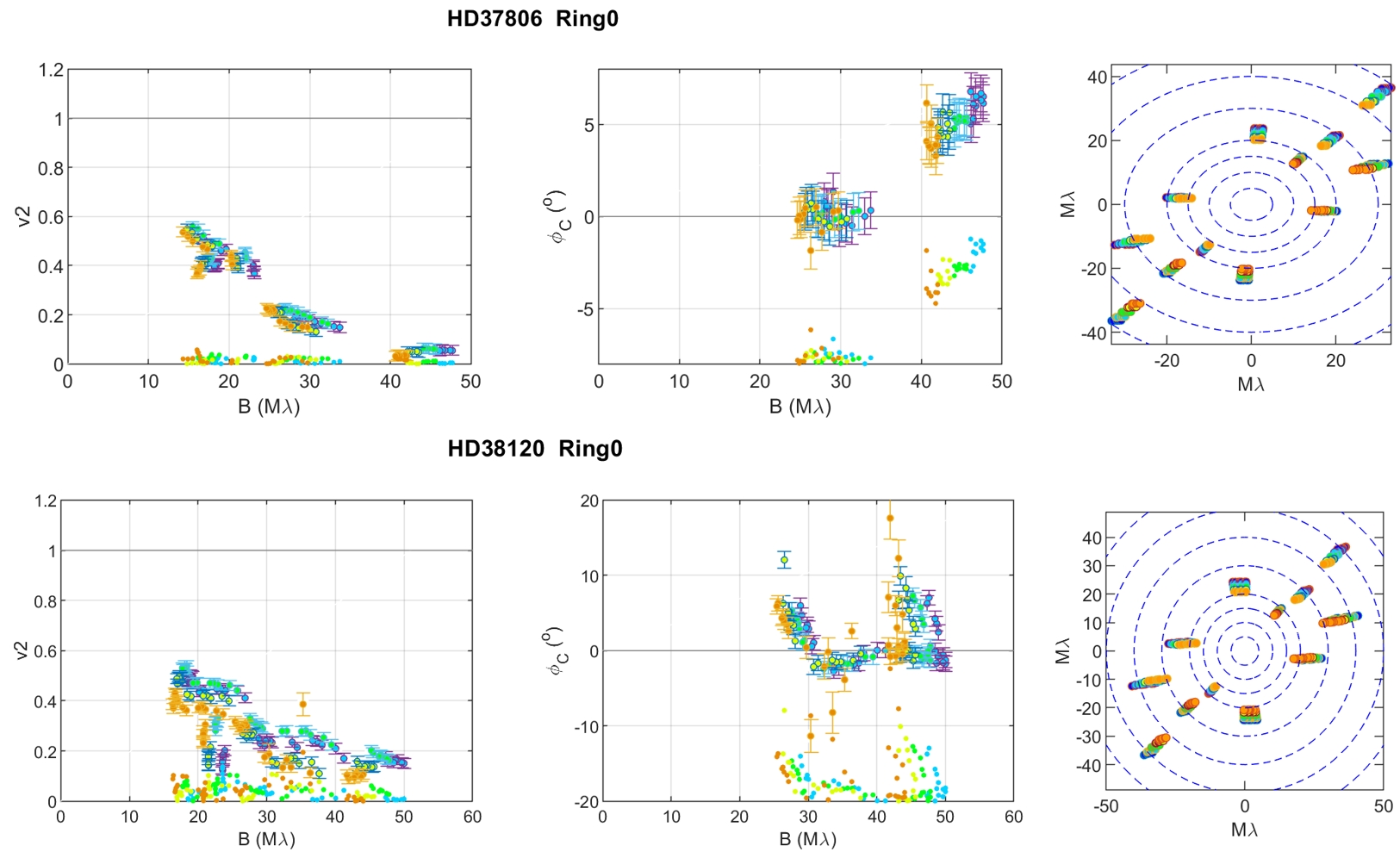}
        \includegraphics[width=15cm]{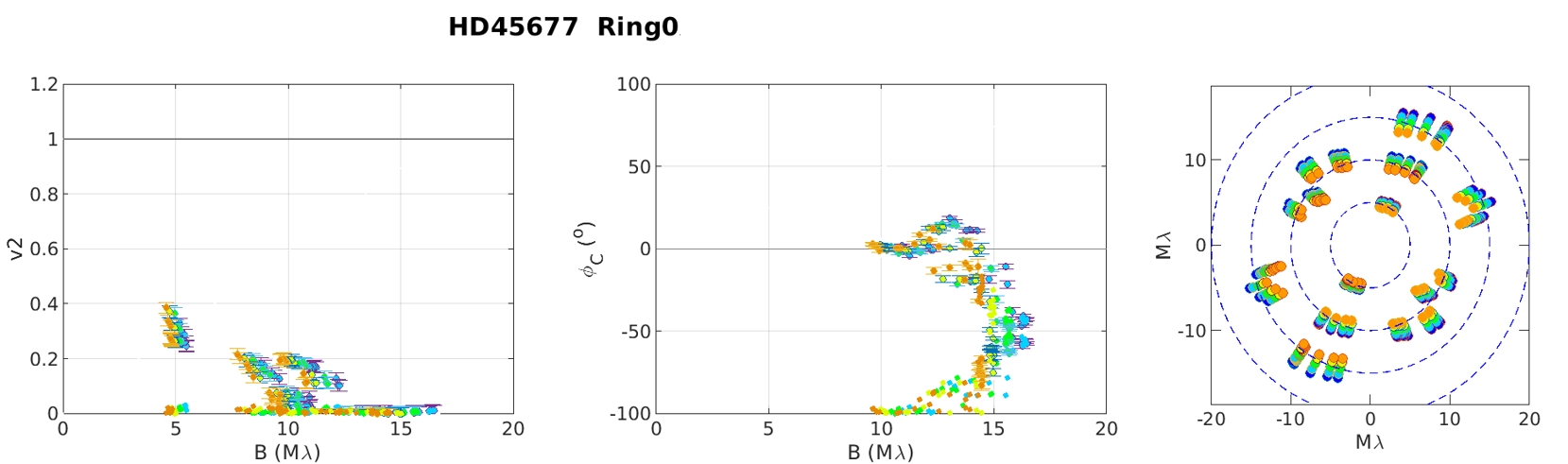}
        \includegraphics[width=15cm]{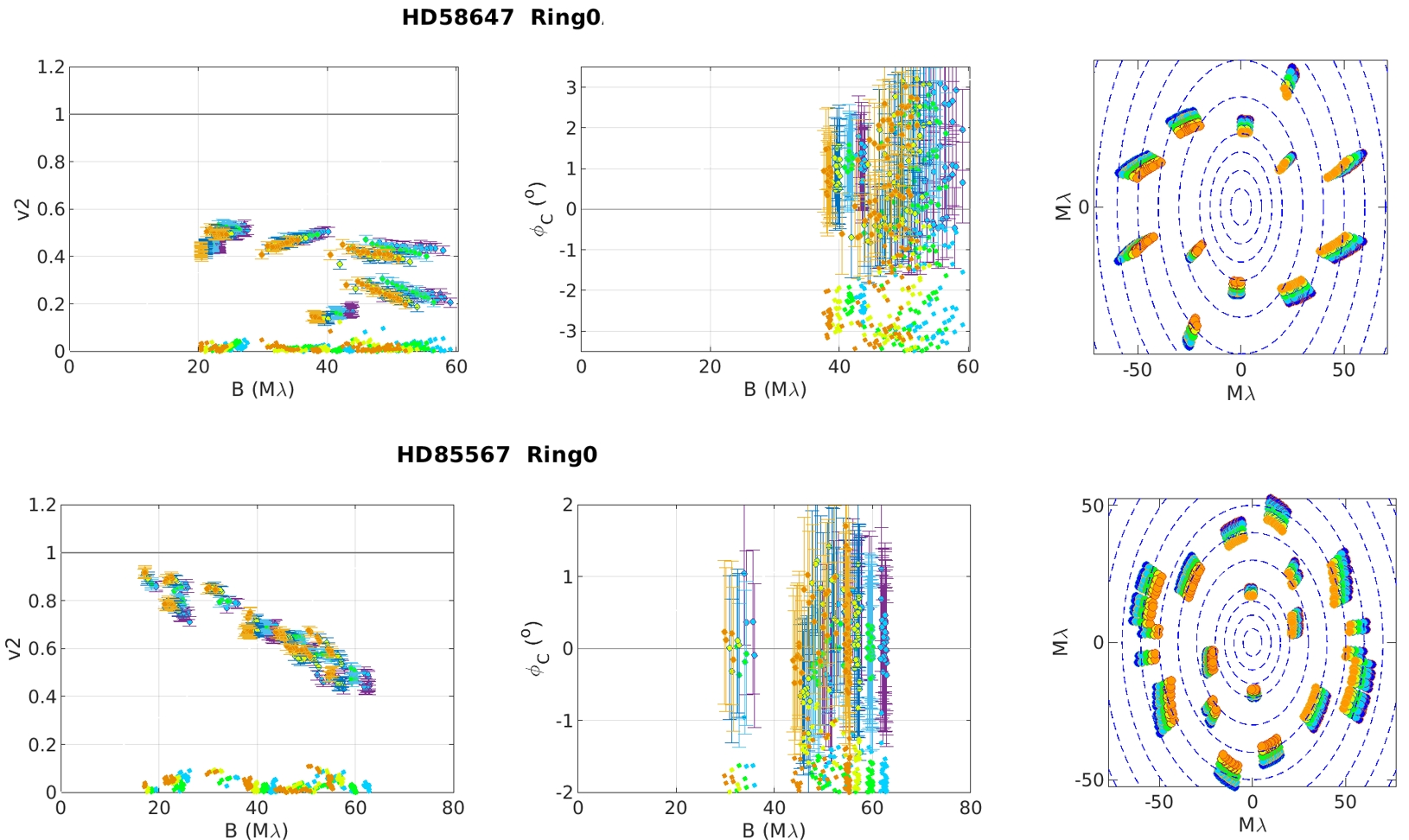}
        \caption{GRAVITY observations: visibilities squared (left), closure phases (middle), (u, v) plane (right). The colors code the spectral channel of the FT. The symbols without contours at the bottom of the visibility and closure phase curves display the residues of the best-fit ring models without azimuthal modulation.}
\end{figure*}
\begin{figure*}[h]
        \centering
        \includegraphics[width=15cm]{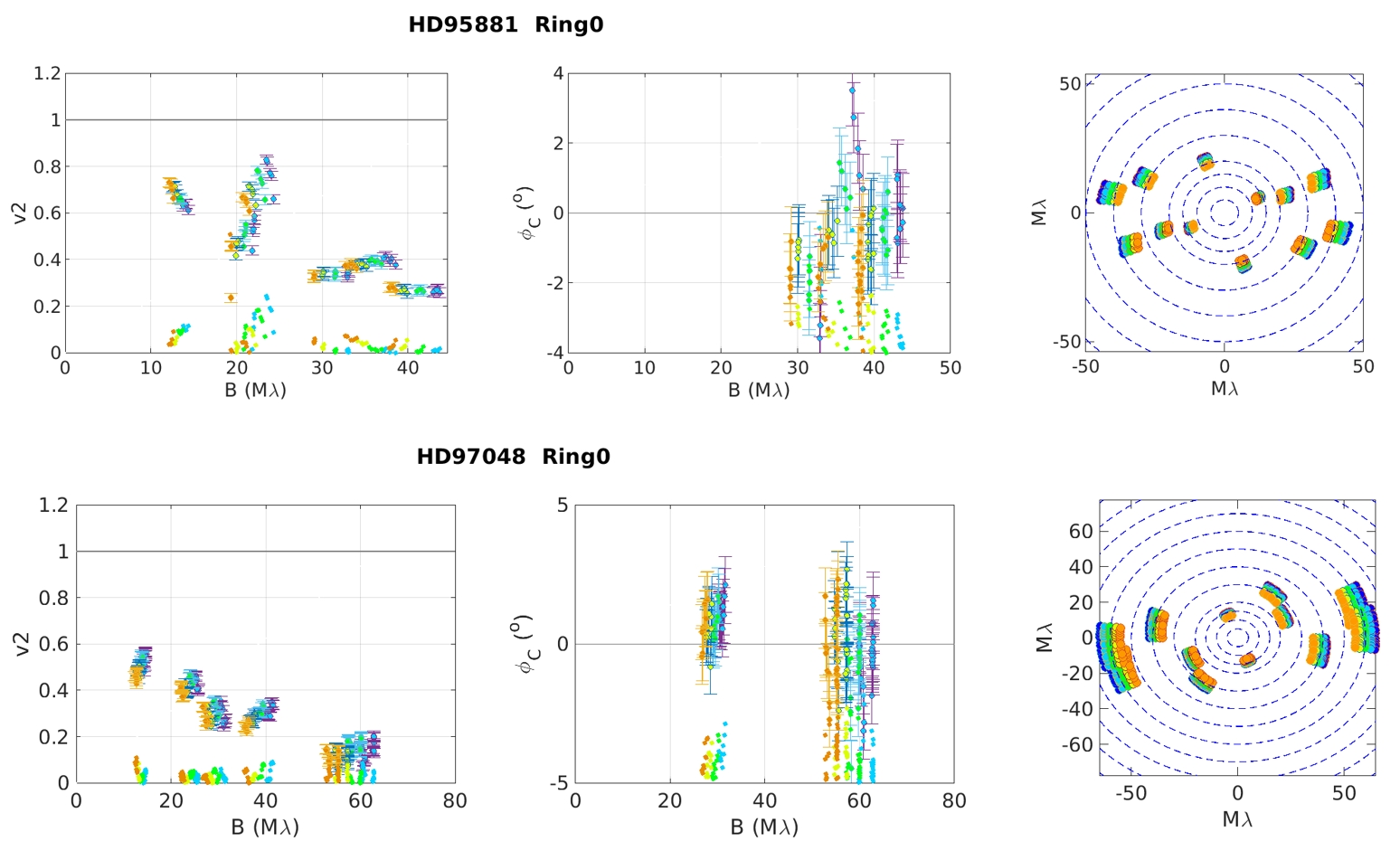}
        \includegraphics[width=15cm]{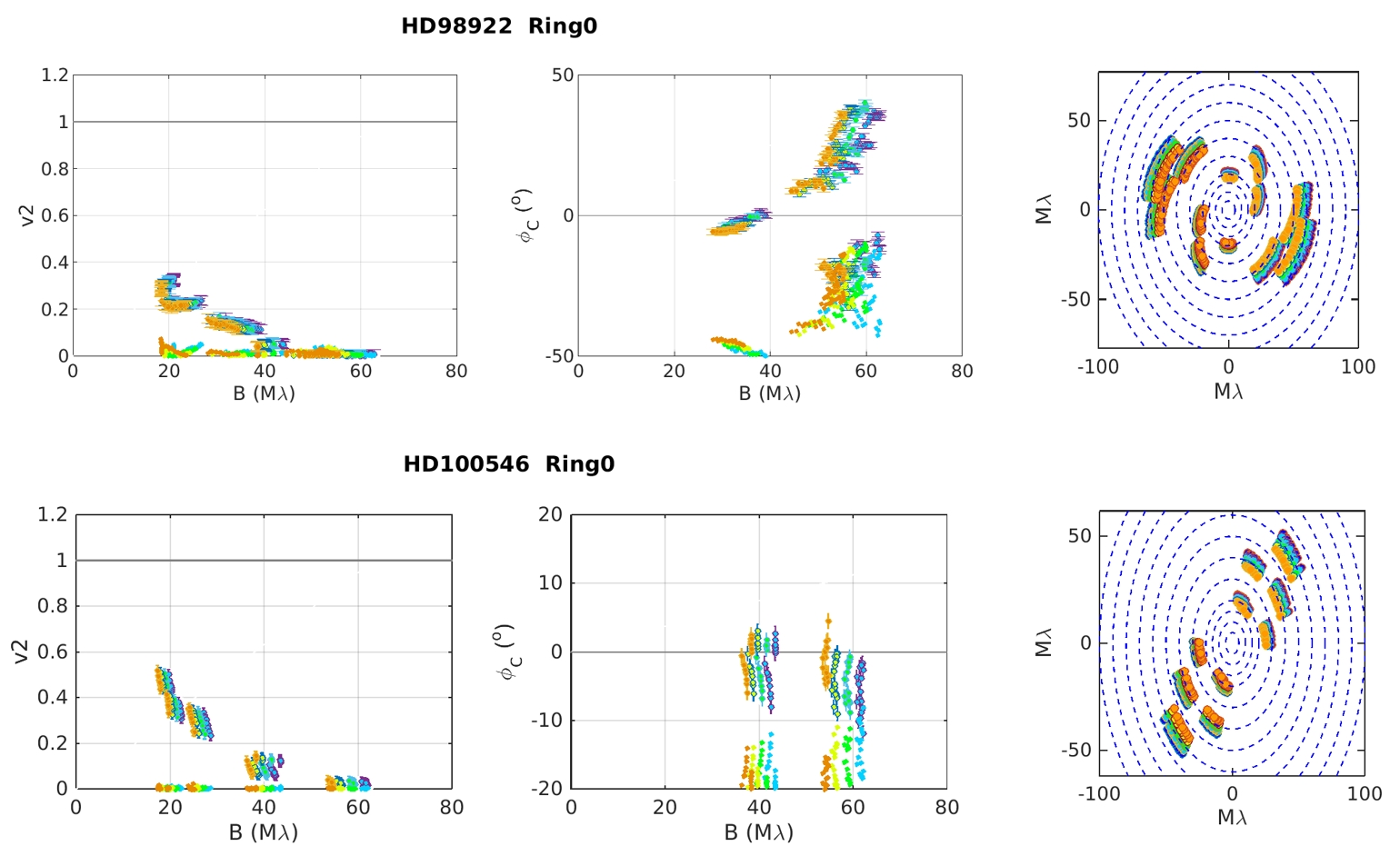}
        \includegraphics[width=15cm]{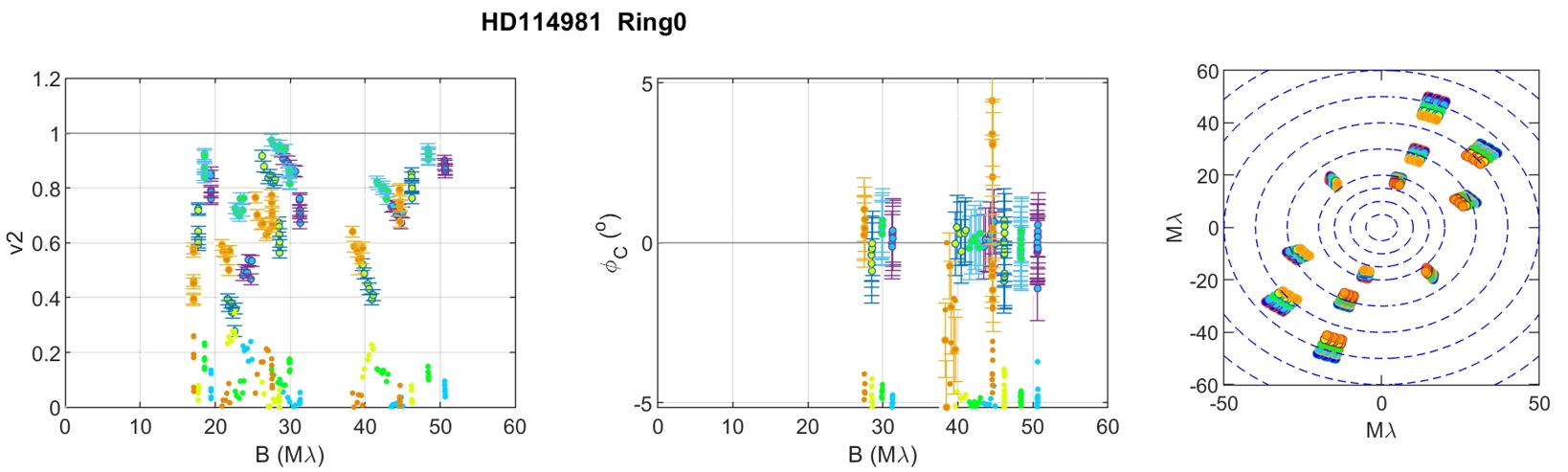}
        \caption{GRAVITY observations (continued)}
        \end{figure*}
\begin{figure*}[h]
        \centering
        \includegraphics[width=15cm]{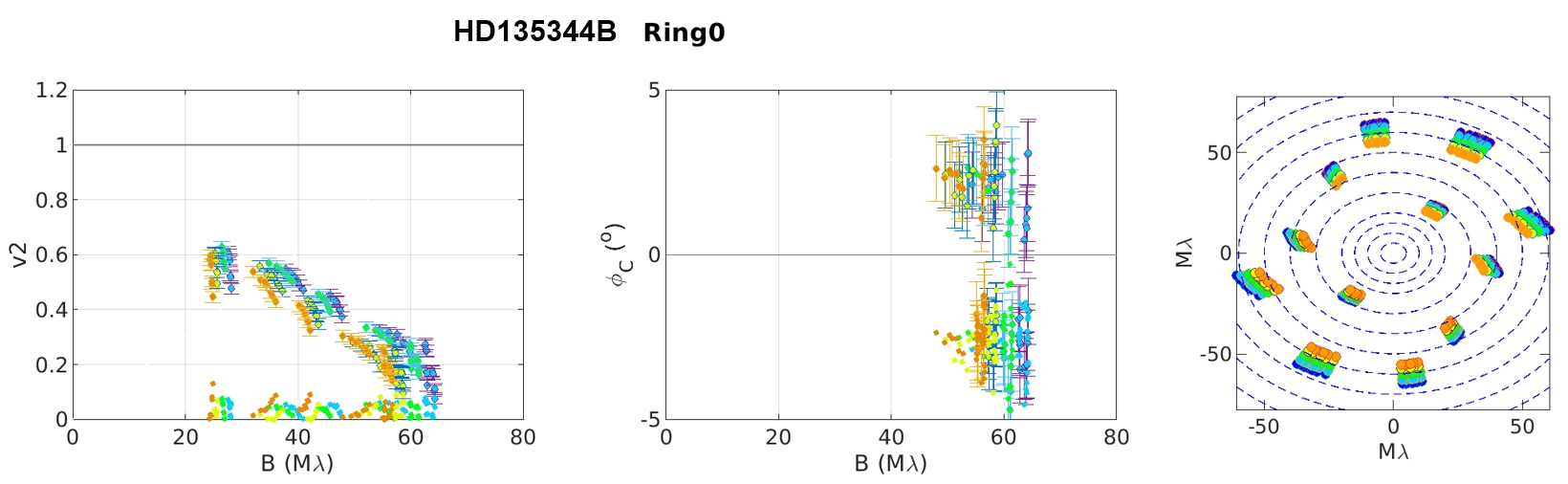}
        \includegraphics[width=15cm]{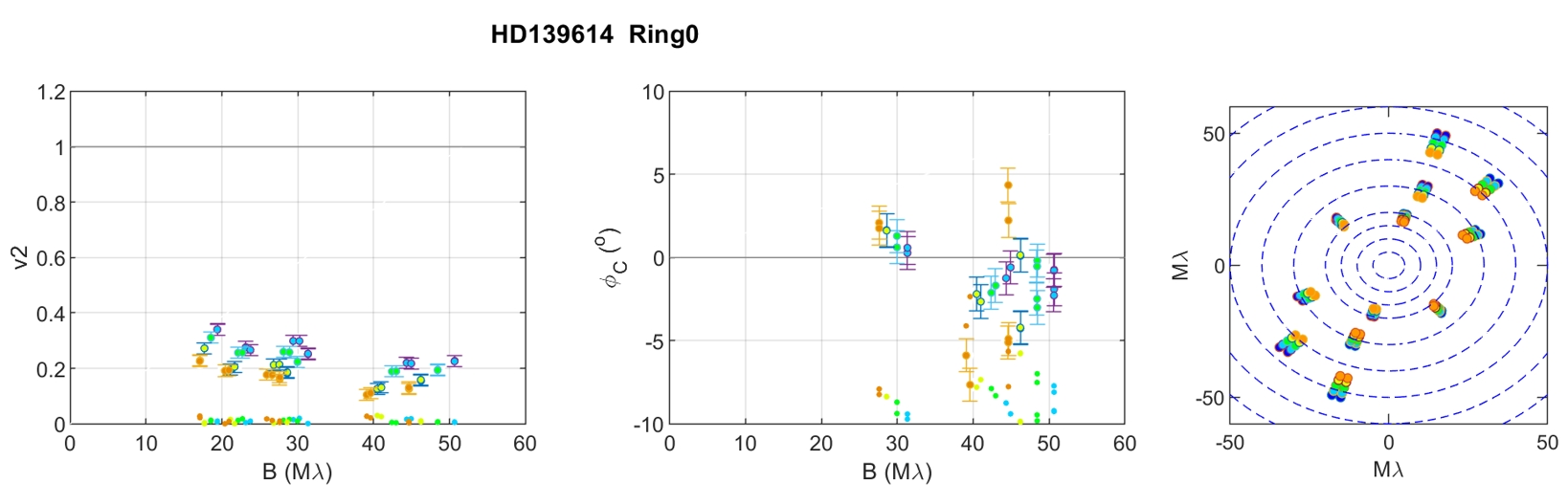}
        \includegraphics[width=15cm]{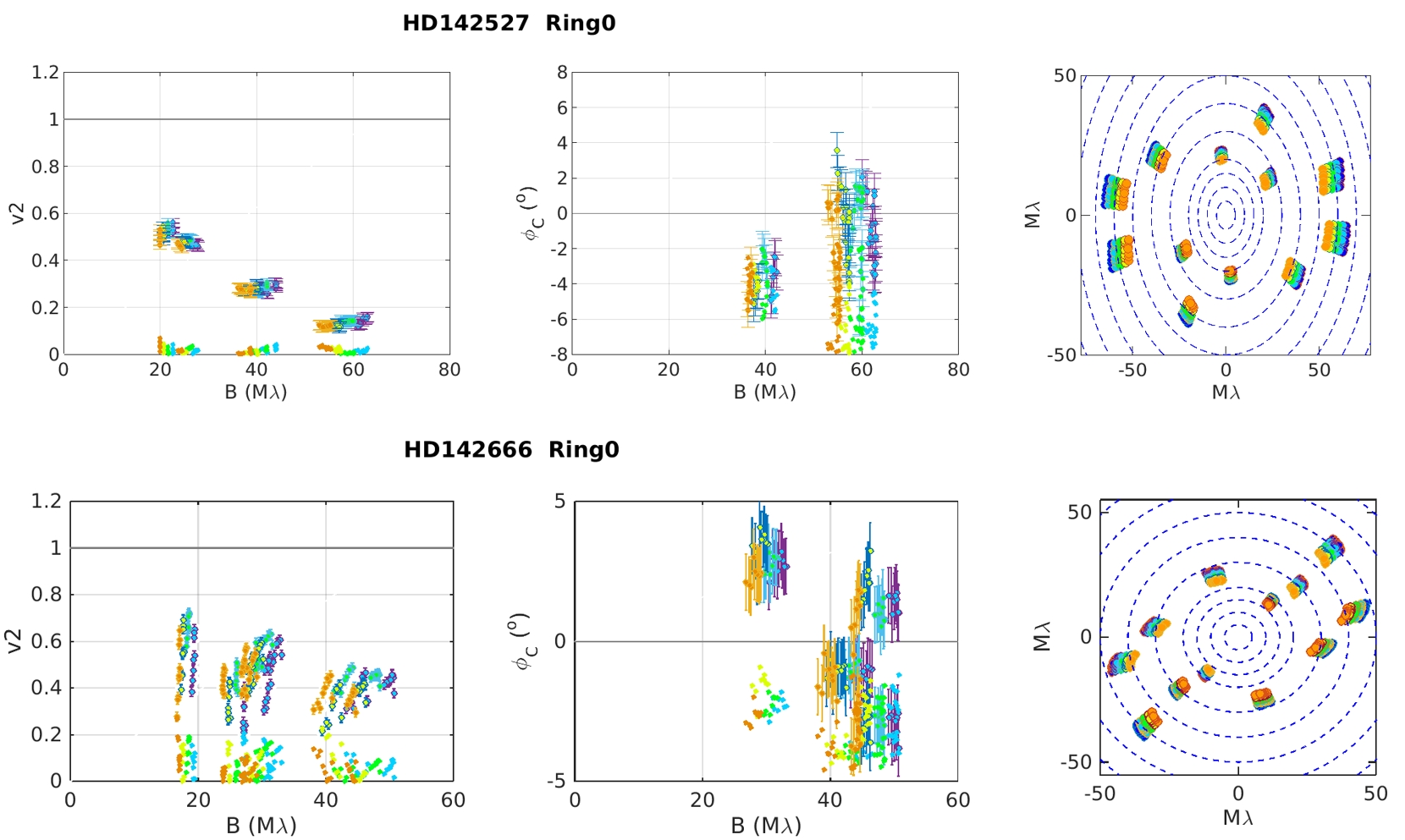}
        \includegraphics[width=15cm]{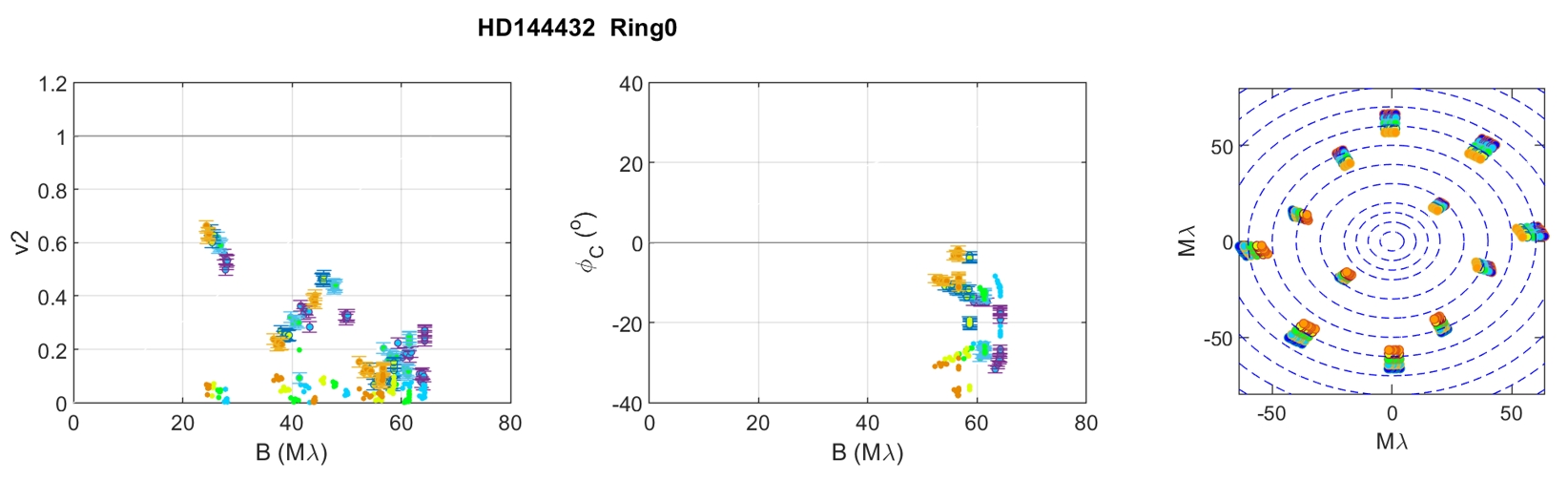}
        \caption{GRAVITY observations (continued)}
\end{figure*}
\begin{figure*}[h]
        \centering
        \includegraphics[width=15cm]{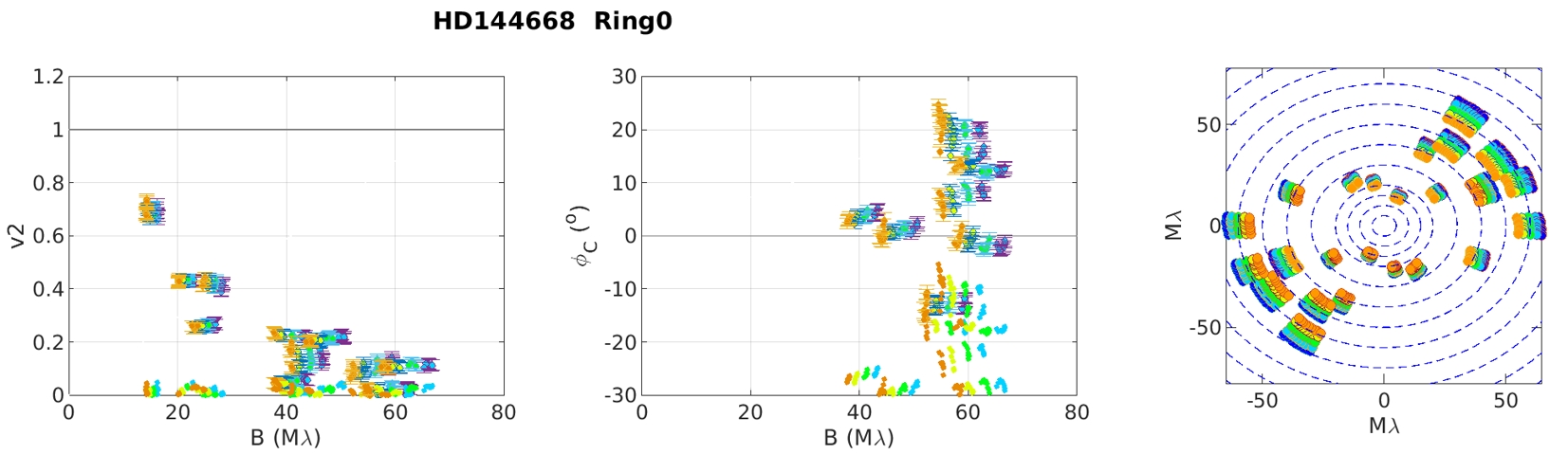}
        \includegraphics[width=15cm]{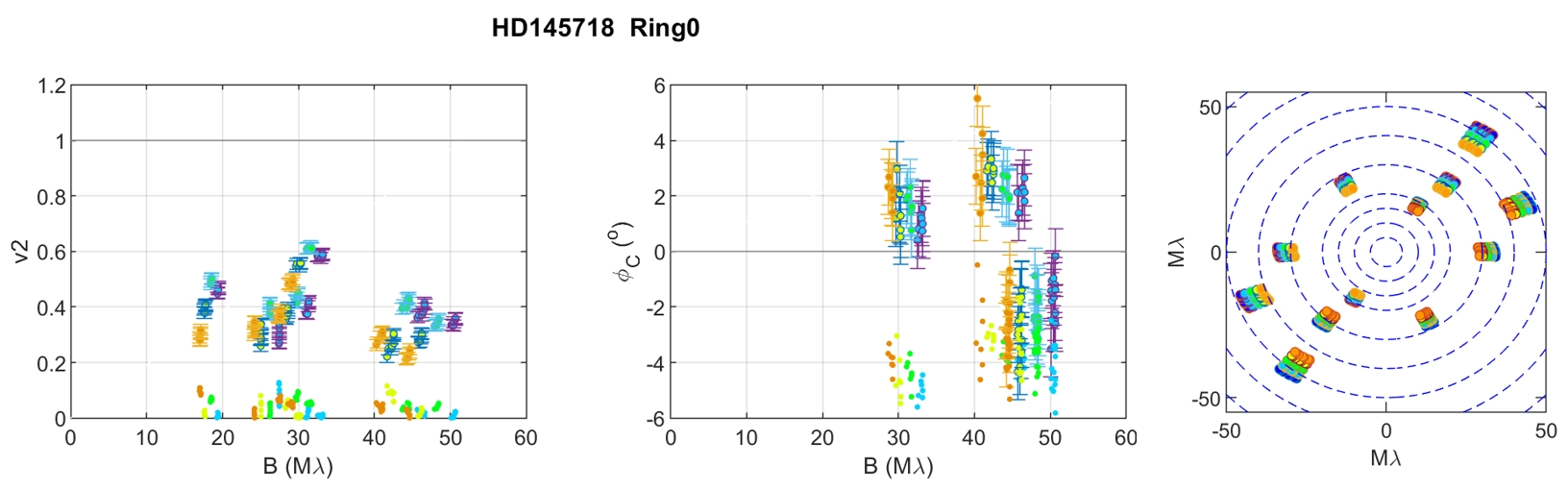}
        \includegraphics[width=15cm]{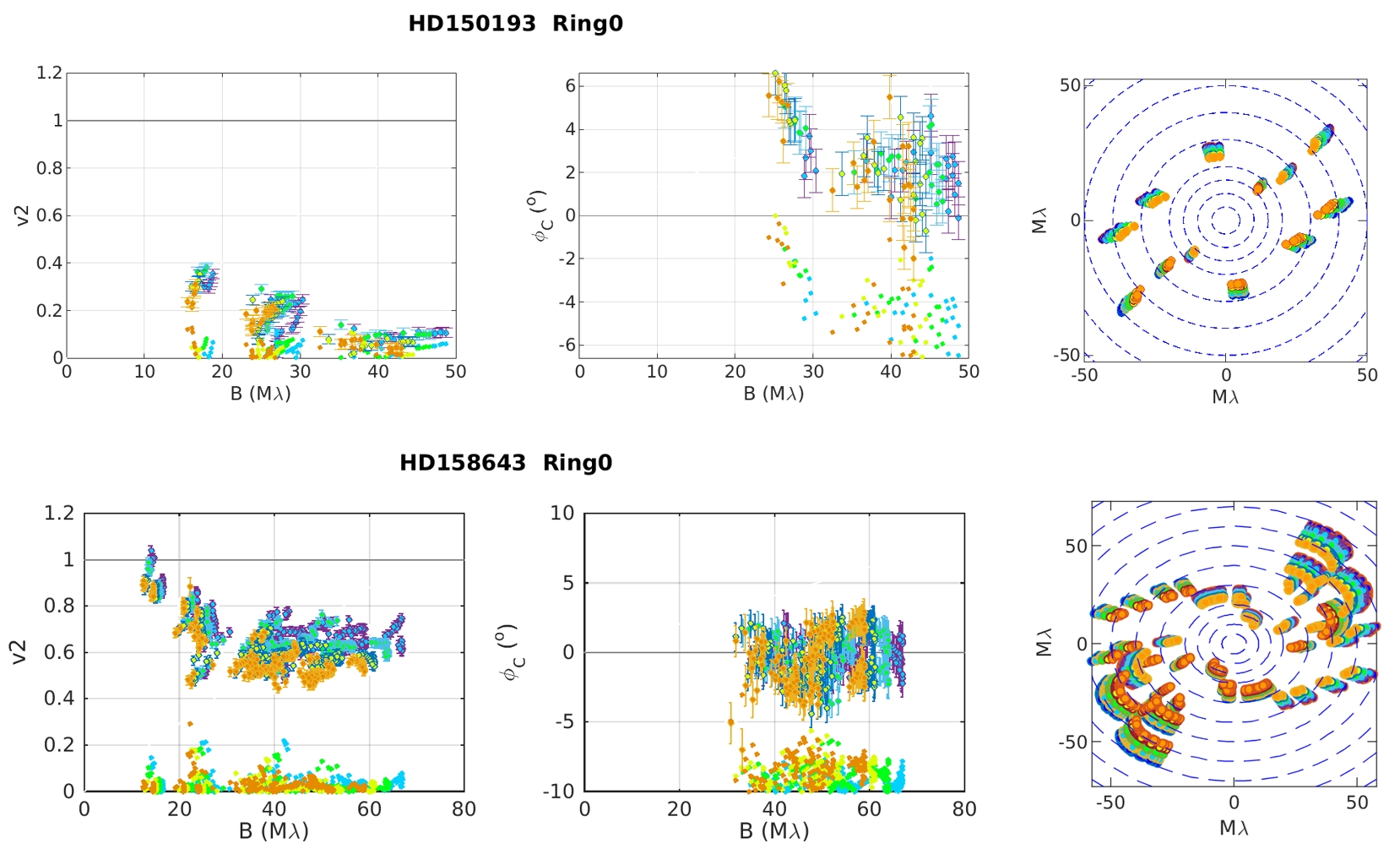}
        \includegraphics[width=15cm]{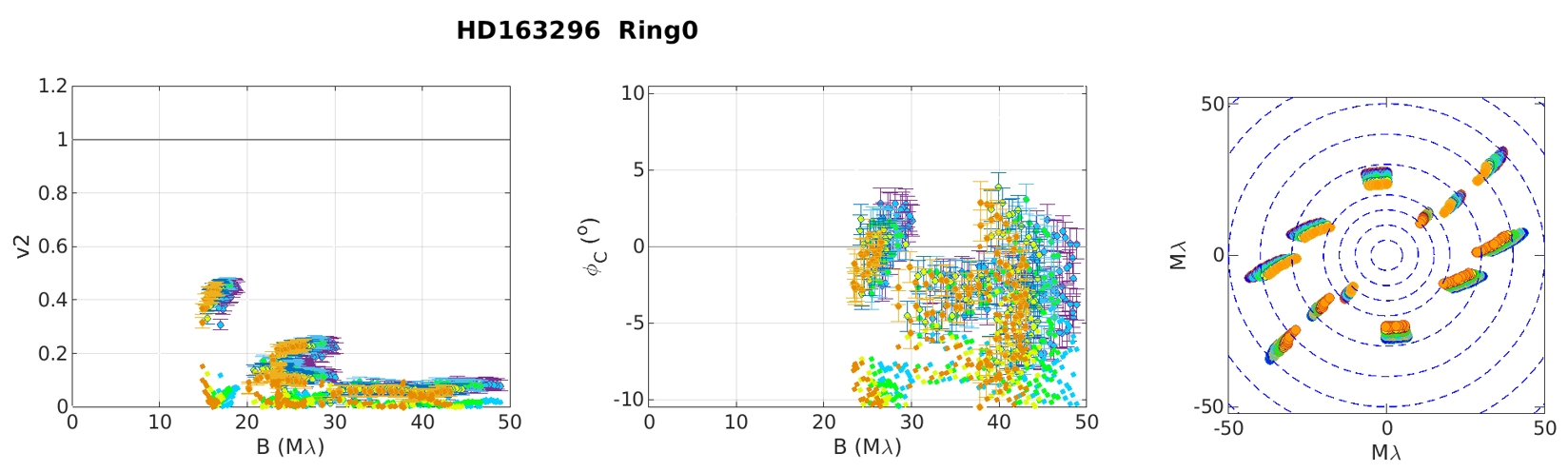}
        \caption{GRAVITY observations (continued)}
\end{figure*}
\begin{figure*}[h]
        \centering
        \includegraphics[width=15cm]{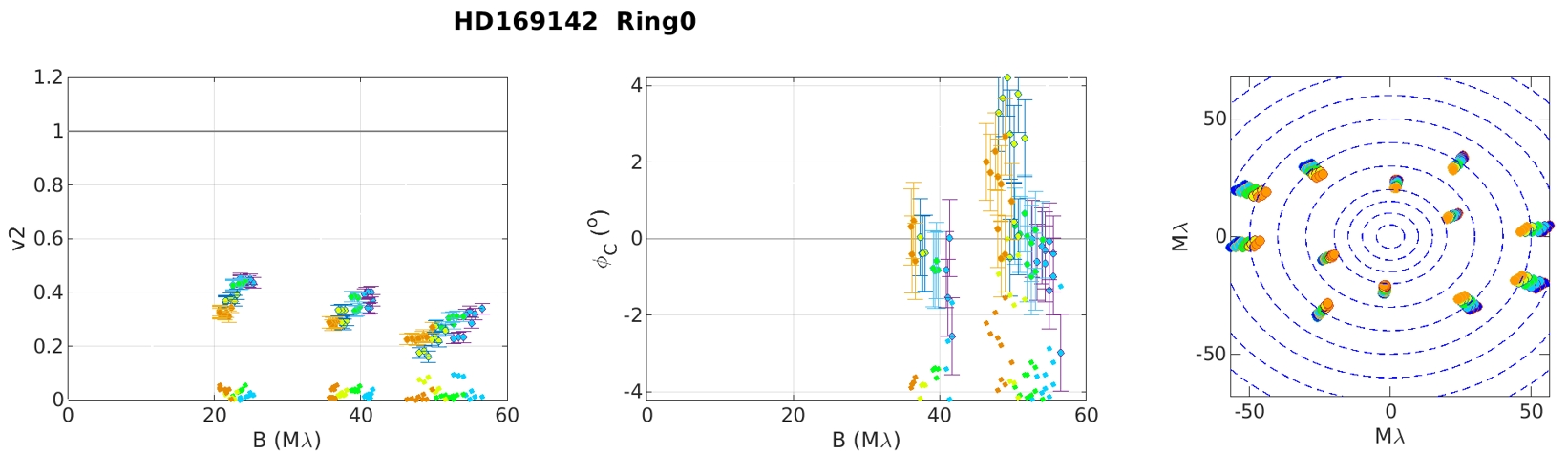}
        \includegraphics[width=15cm]{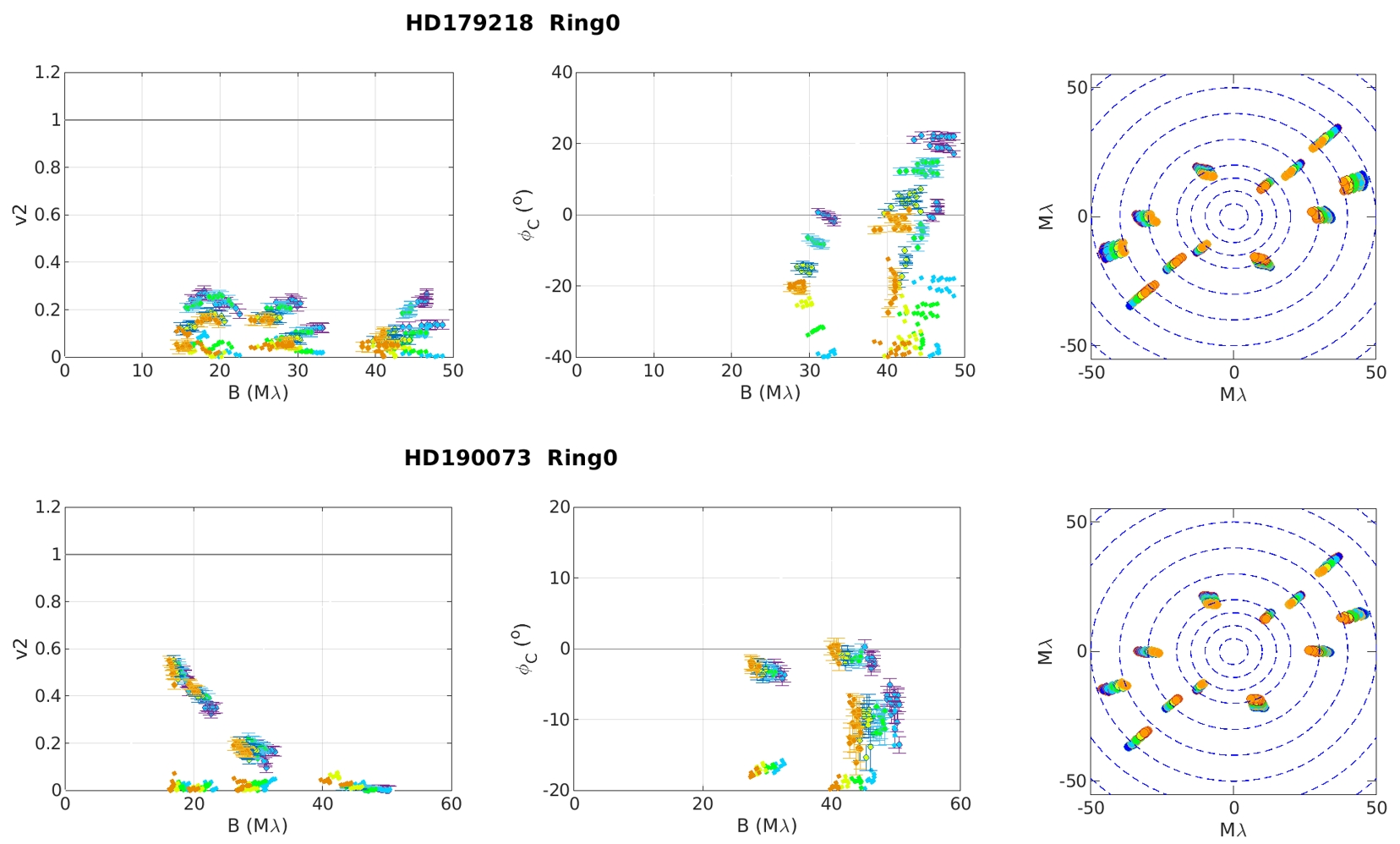}
        \includegraphics[width=15cm]{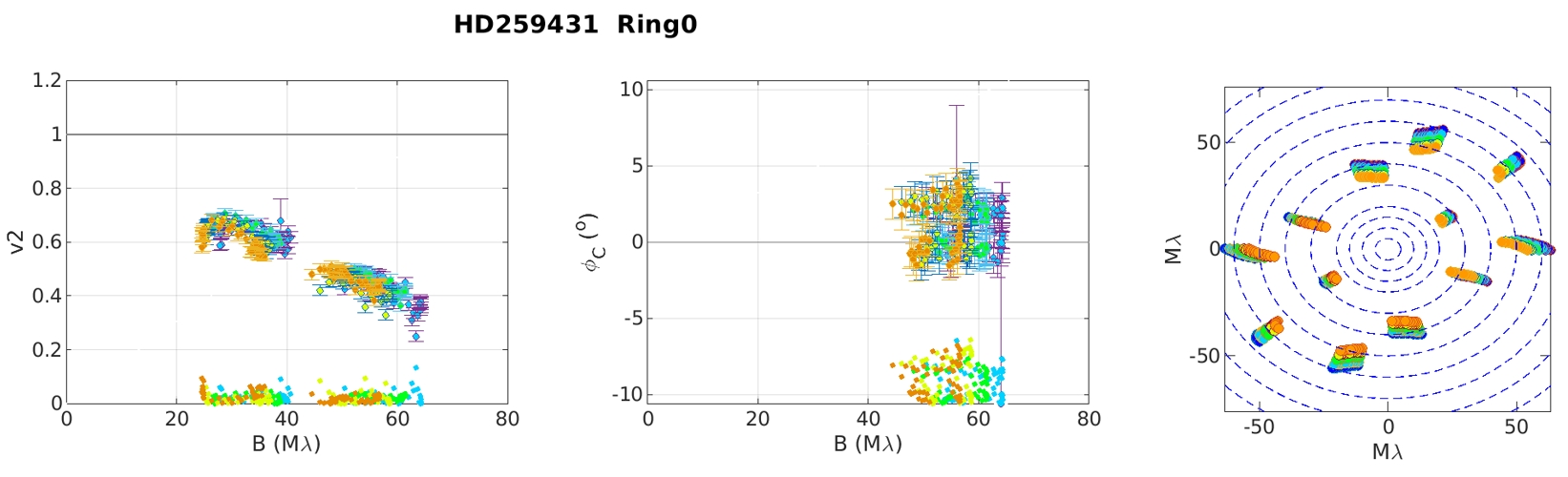}
        \includegraphics[width=15cm]{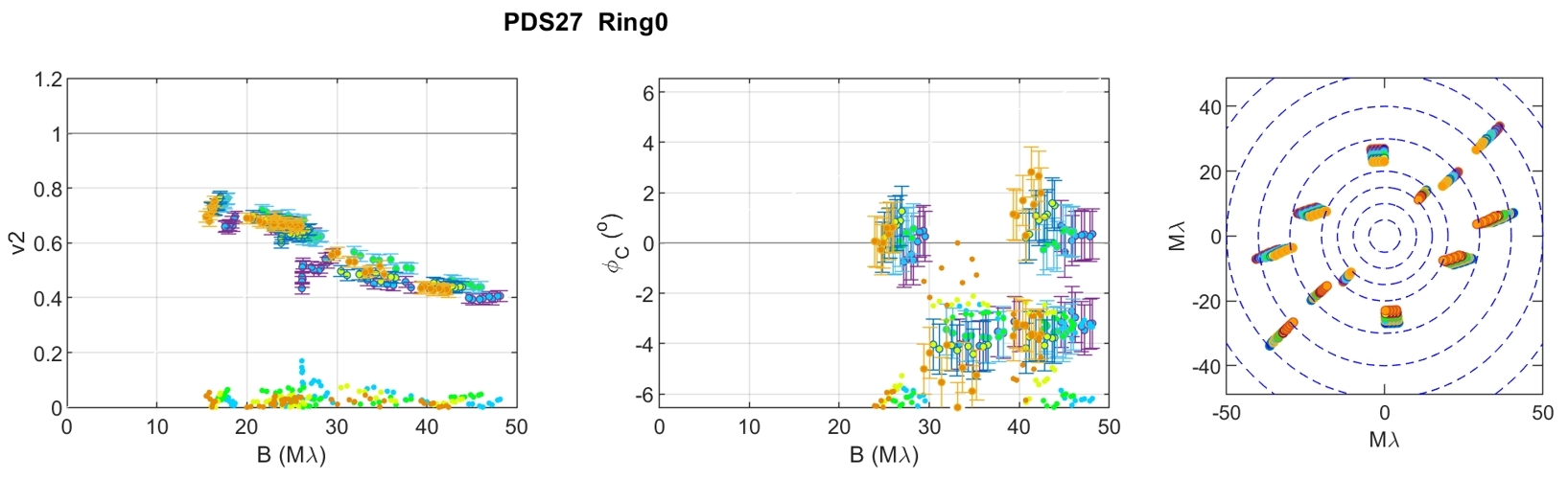}
        \caption{GRAVITY observations (continued)}
\end{figure*}
\begin{figure*}[h]
        \centering
        \includegraphics[width=15cm]{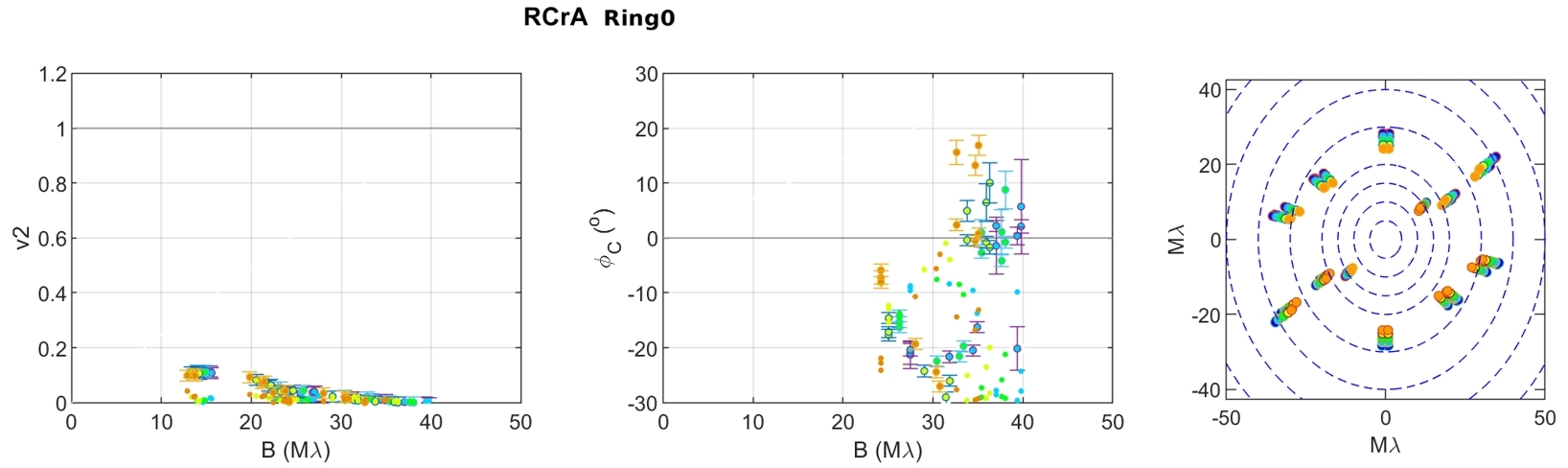}
        \includegraphics[width=15cm]{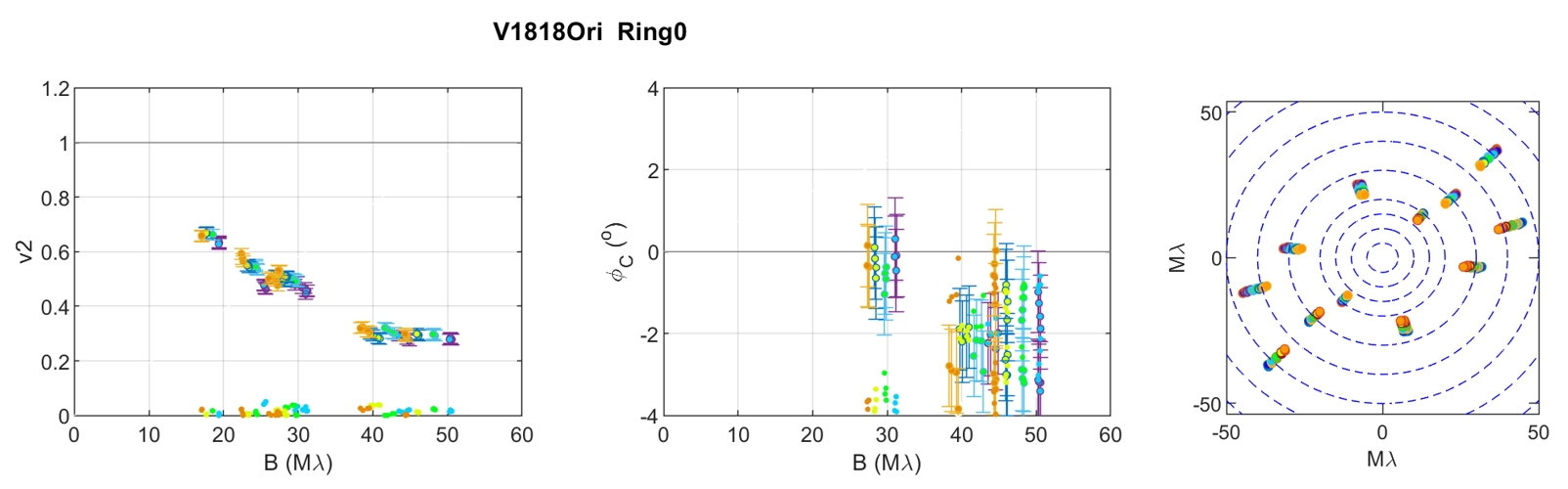}
        \caption{GRAVITY observations (continued)}
    \label{fig:Data}
\end{figure*}

\section{Best-fit Gaussian models}

Results of  the best-fit for a Gaussian ellipsoid are summarised in Table~C.1.

\begin{table*}[h]
\centering
\label{tab:ellip}
\caption{Best-fit parameters for  Gaussian ellipsoid models. Closure phases are not included in the $\chi_{r}^2$ value. Notations are the same as Table 3.}
\vspace{0.1cm}
\begin{tabular}{c c c c c c c c c}
\hline \hline
\# & Object & $f_c$ & $f_h$ & $\cos i$ & $PA$  & FWHM & $k_c$ & $\chi_r^2$\\
& -- & -- & -- & -- & [$^\circ$] & [mas] & -- & \\
\hline 
1 & HD~37806 & 0.82~$\pm$~0.01 & 0.18$\pm$~0.01 & 0.50$\pm$~0.05 & 51~$\pm$~1 & 3.17~$\pm$~0.07 & 2280~$\pm$~310 & 2.25\\ 
2 & HD~38120 & 0.65~$\pm$~0.01 & 0.00$\pm$~0.01 & 0.66$\pm$~0.02 & 164~$\pm$~2 & 6.18~$\pm$~0.14 & 1440~$\pm$~80 & 5.39\\
3 & HD~45677 & 0.89~$\pm$~0.01 & 0.01~$\pm$~0.01 & 0.49~$\pm$~0.01 & 64~$\pm$~1 & 21.4~$\pm$~0.50 & 970~$\pm$~50 & 0.60\\
4 & HD~58647 & 0.64~$\pm$~0.01 & 0.03~$\pm$~0.01 & 0.44~$\pm$~0.01 & 15~$\pm$~1 & 4.18~$\pm$~0.10 & 1070~$\pm$~40 & 1.11\\
5 & HD~85567 & 0.88~$\pm$~0.02 & 0.07~$\pm$~0.02 & 0.94~$\pm$~0.01 & 96~$\pm$~7 & 1.07~$\pm$~0.03 & 1800~$\pm$~200 & 3.03\\
6 & HD~95881 & 0.87~$\pm$~0.01 & 0.09~$\pm$~0.01 & 0.61~$\pm$~0.03 & 167~$\pm$~3 & 3.03~$\pm$~0.14 & 1660~$\pm$~70 & 16.2\\
7 & HD~97048 & 0.74~$\pm$~0.02 & 0.25~$\pm$~0.01 & 0.72~$\pm$~0.02 & 9~$\pm$~3 & 2.14~$\pm$~0.11 & -- & 3.51\\
8 & HD~98922 & 0.85~$\pm$~0.01 & 0.01~$\pm$~0.01 & 0.66~$\pm$~0.01 & 130~$\pm$~1 & 5.90~$\pm$~0.14 & 1320~$\pm$~50 & 1.58\\
9 & HD~100546 & 0.76~$\pm$~0.01 & 0.07~$\pm$~0.01 & 0.65~$\pm$~0.01 & 147~$\pm$~1 & 4.91~$\pm$~0.11 & 1330~$\pm$~90 & 0.64\\
10 & HD~114981 & 0.17~$\pm$~0.01 & 0.00$\pm$~0.01 & 0.13$\pm$~0.07 & 156~$\pm$~5 & 21.9~$\pm$~9.0 & 1290~$\pm$~110 & 38.0\\
11 & HD 135344B & 0.57~$\pm$~0.01 & 0.05$\pm$~0.01 & 0.73$\pm$~0.02 & 31~$\pm$~2 & 2.82~$\pm$~0.14 & 1650~$\pm$~70 & 3.18\\
12 & HD~139614 & 0.59~$\pm$~0.01 & 0.00$\pm$~0.00 & 0.71$\pm$~0.04 & 167~$\pm$~5 & 7.96~$\pm$~0.19 &  1190~$\pm$~60 & 0.73\\
13 & HD~142527 & 0.62~$\pm$~0.02 & 0.13~$\pm$~0.01 & 0.90~$\pm$~0.01 & 16~$\pm$~4 & 2.64~$\pm$~0.06 & 1560~$\pm$~0.21 & 1.67\\
14 & HD~142666 & 0.74~$\pm$~0.01 & 0.24~$\pm$~0.01 & 0.55~$\pm$~0.04 & 155~$\pm$~3 & 1.78~$\pm$~0.12 & 1760~$\pm$~130 & 19.5\\
15 & HD~144432 & 0.73~$\pm$~0.01 & 0.00~$\pm$~0.00 & 0.68~$\pm$~0.02 & 76~$\pm$~2 & 2.89~$\pm$~0.07 & 1600~$\pm$~100 & 6.87\\
16 & HD~144668 & 0.75~$\pm$~0.01 & 0.08~$\pm$~0.01 & 0.56~$\pm$~0.01 & 124~$\pm$~1 & 4.18~$\pm$~0.10 & 1720~$\pm$~80 & 1.58\\
17 & HD~145718 & 0.44~$\pm$~0.01 & 0.00$\pm$~0.00 & 0.37$\pm$~0.01 & 0~$\pm$~2 & 8.93~$\pm$~0.42 & 1490~$\pm$~60 & 5.61\\
18 & HD~150193 & 0.85~$\pm$~0.01 & 0.00~$\pm$~0.00 & 0.68~$\pm$~0.03 & 6~$\pm$~3 & 5.64~$\pm$~0.13 & 1990~$\pm$~80 & 6.85\\
19 & HD~158643 & 0.19~$\pm$~0.01 & 0.04~$\pm$~0.01 & 0.46~$\pm$~0.01 & 116~$\pm$~1 & 5.90~$\pm$~0.14 & 990~$\pm$~20 & 1.61\\
20 & HD~163296 & 0.75~$\pm$~0.01 & 0.01~$\pm$~0.01 & 0.76~$\pm$~0.01 & 136~$\pm$~2 & 6.32~$\pm$~0.15 & 1440~$\pm$~50 & 2.82\\
21 & HD~169142 & 0.37~$\pm$~0.05 & 0.11~$\pm$~0.05 & 0.65~$\pm$~0.1 & 35~$\pm$~5 & 6.1~$\pm$~0.8 & 1130~$\pm$~90 & 2.8\\
22 & HD~179218 & 0.63~$\pm$~0.01 & 0.00~$\pm$~0.00 & 0.59~$\pm$~0.13 & 49~$\pm$~9 & 13.5~$\pm$~3.9 & 1180~$\pm$~50 & 3.92\\
23 & HD~190073 & 0.86~$\pm$~0.01 & 0.04~$\pm$~0.01 & 0.93~$\pm$~0.01 & 64~$\pm$~9 & 3.81~$\pm$~0.09 & 1510~$\pm$~80 & 2.98\\
24 & HD~259431 & 0.85~$\pm$~0.01 & 0.14~$\pm$~0.01 & 0.88~$\pm$~0.02 & 50~$\pm$~3 & 1.02~$\pm$~0.03 & 1530~$\pm$~70 & 2.29\\
25 & PDS 27 & 0.80~$\pm$~0.02 & 0.11$\pm$~0.02 & 0.93$\pm$~0.04 & 155~$\pm$~16 & 1.64~$\pm$~0.10 & 1720~$\pm$~90 & 3.47\\
26 & R CrA & 0.98~$\pm$~0.01 & 0.00~$\pm$~0.00 & 0.52~$\pm$~0.04 & 16~$\pm$~3 & 9.6~$\pm$~0.2 & 1340~$\pm$~90 & 1.20\\
27 & V1818 Ori & 0.63~$\pm$~0.08 & 0.12$\pm$~0.02 & 0.87$\pm$~0.01 & 132~$\pm$~2 & 2.46~$\pm$~0.24 & 1370~$\pm$~110 & 0.83\\
\hline
\end{tabular}
\end{table*}

\section{Visibility of a circumstellar environment with Gaussian and Lorentzian radial brightness distributions }

Our visibility modeling allows us to describe radial brightness distributions from Gaussian one to pseudo-Lorentzian ones through the $\rm Lor$ parameter. $\rm Lor$ ranges between 0 and 1; $\rm Lor$~=~0 for a Gaussian radial distribution and $\rm Lor$~=~1 for a Lorentzian one. The visibility of the circumstellar environment is thus described by
\begin{equation}
    V_c = (1 - Lor) V_{Gauss} + Lor V_{Lor},
\end{equation}
where $V_{Gauss}$ is the visibility of a pure Gaussian radial brightness distribution and $V_{Lor}$ is the visibility of a pure Lorentzian radial brightness distribution. In both cases, there is an analytical expression of the visibility as given in Table 5 of L17.

\section{Half-flux radii in H- and K-bands}

For the targets that have been observed with PIONIER and GRAVITY, we use the ring half-flux radii in H-band published in L17 to compare them with our GRAVITY measurements in the K-band (Table 3). All radii in mas are converted in au with  distances derived from GAIA DR2 (see Table 1). All values are gathered in Table E.1.

\begin{table}[h]
\centering
\label{tab:HvsK}
\caption{Half-flux radii of  ring models in H- and K-bands with their 1-$\sigma$ error bars.}
\vspace{0.1cm}
\begin{tabular}{c c c }
\hline \hline
Object & $R$ in H-band & $R$ in K-band\\
 -- & [au] & [au] \\
\hline 
HD~37806 & 0.71~$\pm$~0.04 & 0.87~$\pm$~0.05 \\ 
HD~45677 & 5.9~$\pm$~0.4 & 5.9~$\pm$~0.4\\
HD~58647 & 0.57~$\pm$~0.04 & 0.64~$\pm$~0.02\\
HD~85567 & 0.62~$\pm$~0.05 & 0.60~$\pm$~0.03\\
HD~95881 & 1.3~$\pm$~0.1 & 1.7~$\pm$~0.2\\
HD~97048 & 0.21~$\pm$~0.02 & 0.41~$\pm$~0.01\\
HD~98922 & 1.23~$\pm$~0.06 & 1.62~$\pm$~0.08\\
HD~100546 & 0.289~$\pm$~0.007 & 0.276~$\pm$~0.007\\
HD~139614 & 0.19~$\pm$~0.01 & 0.65~$\pm$~0.03 \\
HD~142527 & 0.189~$\pm$~0.009 & 0.198~$\pm$~0.005\\
HD~142666 & 0.159~$\pm$~0.008 & 0.11~$\pm$~0.02\\
HD~144432 & 0.191~$\pm$~0.009 & 0.214~$\pm$~0.006\\
HD~144668 & 0.32~$\pm$~0.01 & 0.33~$\pm$~0.01\\
HD~145718 & 0.22~$\pm$~0.03 & 0.7~$\pm$~0.1\\
HD~150193 & 0.251~$\pm$~0.008 & 0.40~$\pm$~0.01\\
HD~158643 & 0.11~$\pm$~0.02 & 0.38~$\pm$~0.03\\
HD~163296 & 0.25~$\pm$~0.01 & 0.30~$\pm$~0.01\\
HD~169142 & 0.11~$\pm$~0.02 & 0.33~$\pm$~0.07\\
HD~190073 & 1.6~$\pm$~0.2 & 1.8~$\pm$~0.2\\
HD~259431 & 0.32~$\pm$~0.03 & 0.36~$\pm$~0.02 \\
R CrA & 0.44~$\pm$~0.06 & 0.53~$\pm$~0.08\\
\hline
\end{tabular}
\end{table}

\end{appendix}

\end{document}